\documentclass{jfm}
\usepackage{graphicx,psfrag}
\usepackage{caption} 
\usepackage{graphicx}
\usepackage{subcaption}
\usepackage{natbib}
\usepackage{float}
\usepackage{rotating}
\usepackage{amsmath}
\usepackage{latexsym}
\usepackage{amssymb}
\usepackage{mathrsfs}
\usepackage{layout}
\usepackage{bm}% bold math
\usepackage{url}
\usepackage{epstopdf}
\usepackage{wrapfig}
\usepackage{tabularx}
\usepackage{fancyhdr}
\usepackage{color}
\usepackage[makeroom]{cancel}

\ifCUPmtlplainloaded \else
  \checkfont{eurm10}
  \iffontfound
    \IfFileExists{upmath.sty}
      {\typeout{^^JFound AMS Euler Roman fonts on the system,
                   using the 'upmath' package.^^J}%
       \usepackage{upmath}}
      {\typeout{^^JFound AMS Euler Roman fonts on the system, but you
                   dont seem to have the}%
       \typeout{'upmath' package installed. JFM.cls can take advantage
                 of these fonts,^^Jif you use 'upmath' package.^^J}%
      }
  \else
  \fi
\fi

\ifCUPmtlplainloaded \else
  \checkfont{msam10}
  \iffontfound
    \IfFileExists{amssymb.sty}
      {\typeout{^^JFound AMS Symbol fonts on the system, using the
                'amssymb' package.^^J}%
       \usepackage{amssymb}%

      }{}
  \fi
\fi

\ifCUPmtlplainloaded \else
  \IfFileExists{amsbsy.sty}
    {\typeout{^^JFound the 'amsbsy' package on the system, using it.^^J}%
     \usepackage{amsbsy}}
    {\providecommand\boldsymbol[1]{\mbox{\boldmath $##1$}}}
\fi

\newcommand{\nid}{\noindent}

\newcommand{\mi}{\mathrm{i}}

\DeclareMathAlphabet\mathsfbi{OT1}{cmss}{m}{sl}

\providecommand\bcdot{\boldsymbol{\cdot}}
\newcommand{\bs}{\boldsymbol}

\usepackage{tikz}
\newcommand*\circled[1]{\tikz[baseline=(char.base)]{
            \node[shape=circle,draw,inner sep=1pt] (char) {#1};}}
            
\title[]{Clustering of Rapidly Settling, Low-Inertia 
Particle Pairs in Isotropic Turbulence. \\ I. Drift and Diffusion Flux Closures}

\author[Sarma L. Rani, Vijay K. Gupta, and Donald L. Koch]%
{S\ls A\ls R\ls M\ls A\ns L.\ns R\ls A\ls N\ls I$^1$%
  \thanks{Email address for correspondence: sarma.rani@uah.edu},
 V\ls I\ls J\ls A\ls Y\ns K.\ns G\ls U\ls P\ls T\ls A$^1$%
 \thanks{Current Address: Department of Chemical Engineering, 
 University of Missouri, Columbia, Missouri 65211, U.S.A.}, \and 
\break
D\ls O\ls N\ls A\ls L\ls D\ns L.\ns K\ls O\ls C\ls H$^2$}

\affiliation{$^1$Department of Mechanical and Aerospace Engineering, University of Alabama in Huntsville,
Huntsville, Alabama 35899, U.S.A.\\[\affilskip]
$^2$Smith School of Chemical and Biomolecular Engineering, Cornell University, Ithaca, New York 14853, U.S.A.}

\begin{document}

\maketitle
\begin{abstract}

In this two--part study, we present the development and analysis of a 
stochastic theory for characterizing the relative positions 
of monodisperse, low-inertia particle pairs that are 
settling rapidly in homogeneous isotropic turbulence.  
In the limits of small Stokes number and 
Froude number such that $Fr \ll St_\eta \ll 1$,     
%Here $St_\eta$~is the ratio of 
%the  particle viscous relaxation time to the Kolmogorov time scale, and 
%$Fr$ is the ratio of the Kolmogorov scale of 
%acceleration and the magnitude of gravitational acceleration.  
%In these parametric limits, 
closures are developed for the drift and diffusion 
fluxes in the probability density function (PDF) equation for the 
pair relative positions. The theory focuses on the relative motion 
of particle pairs in the dissipation regime of turbulence, i.e., for  
pair separations smaller than the Kolmogorov length scale.  
In this regime, the theory approximates the fluid velocity field 
in a reference frame following the primary particle as locally linear.

In this Part I paper, we present the derivation of closure approximations  
for the drift and diffusion fluxes in the PDF equation for the relative positions $\bs{r}$. 
The drift flux contains the time integral of the third and fourth moments of the ``seen" fluid  
velocity gradients along the trajectories of primary particles. 
These moments may be analytically resolved by making approximations  
regarding the ``seen" velocity gradient.  Accordingly, two 
closure forms are derived specifically for the drift flux.  The first invokes the 
assumption that the fluid velocity gradient along particle trajectories 
has a Gaussian distribution.  In the second drift closure, we instead assume that 
the ``seen" strain-rate and rotation-rate tensors 
scaled by the turbulent dissipation rate and enstrophy, respectively, are 
normally distributed.   A key feature of the second closure is that it 
accounts for the two-time autocorrelations and cross-correlations of dissipation rate 
and enstrophy.  These correlations quantify, as well as illustrate  
the mechanisms driving particle clustering.  
%Furthermore, particle clustering 
%predicted through the use of the first drift closure, while influenced by 
%gravity, is independent of $Fr$.  However, the second closure retains an 
%explicit dependence on $Fr$.
Analytical solution to the PDF $\langle P \rangle (r,\theta)$ is then derived, where  
the $\theta$ is spherical polar angle.  It is seen that the PDF has a 
power-law dependence on separation~$r$ of the form 
$\langle P \rangle (r,\theta) \sim r^\beta$, with 
$\beta \sim St_\eta ^2$ and $\beta < 0$, analogous to that for 
the radial distribution function of non-settling pairs.  An explicit expression 
is derived for $\beta$ in terms of the drift and diffusion closures. 
The $\langle P \rangle (r,\theta)$ solution 
also shows that for a given $r$, the clustering of $St_\eta \ll 1$ particles 
is only weakly anisotropic, which is in conformity with prior observations from 
direct numerical simulations of isotropic turbulence containing settling particles.

\end{abstract}

\section{Introduction}

This paper presents a stochastic theory for inertial particle clustering that incorporates the 
effects of settling on the sampling of turbulence.  The theory focuses on the  
relative motion of low-Stokes-number pairs for sub-Kolmogorov separations.  The study is principally 
motivated by the desire to understand the microphysical processes influencing the 
relative motion of water droplets in cumulus clouds.  

The growth of droplets in a cloud from a radius of $10$-$20~\mu$m to 
raindrop size ($> 100~\mu$m radius) is a central problem in cloud physics.  
Cloud microphysical models describe droplet growth through two main 
mechanisms: (1) condensation, and (2) droplet collision and coalescence.  
For radii $< 20~\mu$m, droplet growth is principally driven by 
condensation \citep{bartlett1966}.  For larger radii, collision and coalescence play 
an increasingly important role, eventually becoming the dominant mechanism for~radii~$> 40~\mu$m. 
Interestingly, in the $15$-$40~\mu$m radius range, droplet Stokes numbers 
$St_\eta$ are in the $0.1$-$1$ range.  The relative motion of 
such droplet pairs is strongly susceptible to the effects of 
air turbulence.  For instance, it is now well established that for $St_\eta < 1$, particles 
exhibit strong spatial clustering arising from the complex interactions between 
turbulent eddies and particle inertia \citep{chun2005,bragg2014a,bragg2014b}.  
Turbulence-induced clustering of droplets
may lead to increased collision rates, potenially playing a key role in droplet growth.
In addition to turbulence, differential gravitational settling among droplets
is an important driver of collisions, particularly for pairs of 
larger drops whose size ratio departs substantially from one.  
Differential settling also reduces the clustering of particles 
with different radii so that the most pronounced inertial clustering 
occurs in drops of nearly equal size \citep{ayala08a,parishani2015}.

In cumulus and stratocumulus clouds, the Kolmogorov-scale fluid acceleration ($a_\eta$) is small relative 
to gravitational acceleration ($g$) so that the Froude number 
$Fr = a_\eta/g \sim 0.009$-$0.06$ \citep{ayala08b,fouxon2015}.  Therefore, the present study focuses on 
the relative motion of monodisperse, low-inertia particle 
pairs that are undergoing rapid settling in isotropic turbulence.  
While $Fr$ characterizes fluid accelerations, the settling velocity parameter
$Sv_\eta$ is used to quantify particle settling, 
where $Sv_\eta$ is defined as the ratio of particle terminal velocity to the Kolmogorov velocity 
scale.  Therefore, by rapid settling, we mean $Sv_\eta \gg 1$.  Recognizing that 
$Sv_\eta = St_\eta/Fr$, the current stochastic theory is   
derived in the regime characterized by $Fr \ll St_\eta \ll 1$.  
Here Stokes number $St_\eta$ is the ratio 
of the particle viscous relaxation time $\tau_v$ and the Kolmogorov time scale $\tau_\eta$. 
In these parametric limits, the transport equation 
for the probability density function (PDF) of pair separations ($\bs{r}$) is 
of the drift-diffusion form.  In this Part~I paper, we derive closure approximations 
for the drift and diffusion fluxes.  The PDF equation is 
also solved analytically, giving rise to a PDF with a power-law dependence 
on pair separation $r$ with a negative exponent.  An explicit expression is obtained 
for the exponent in terms of the drift and diffusion fluxes.  
%The polar--angle dependence 
%of the PDF shows that the theory predicts the formation of vertical columns 
%of sedimenting particles in isotropic turbulence \citep{bec2014,park2014}.

Turbulence--driven inhomogeneities in the spatial distribution of inertial particles 
are believed to play an important role in locally enhancing particle collision rates.  
Preferential concentration is one of the mechanisms of particle clustering, wherein inertial 
particles denser than the fluid are ejected out of vorticity-dominated regions, and accumulate in  
strain-dominated regions. Numerous computational, experimental and theoretical studies 
of aerosol dynamics in isotropic turbulence have established that inertial 
particles preferentially concentrate in regions of excess strain-rate over rotation-rate \citep{maxey1987, 
squires1991, eaton1994, druzhinin, druzhinin99, rani1, rani2, rani3, chun2005, ray2011}.  

Since the characteristic length scales of strain rate and rotation rate in isotropic 
turbulence scale with the Kolmogorov length scale ($\eta$), it may be expected that 
preferential concentration enhances the probability of finding a pair of particles 
at separations comparable to $\eta$.  However, \citet{reade2000effect} showed 
through direct numerical simulations (DNS) of particle-laden isotropic turbulence that 
inertial particles continued to exhibit clustering at 
separations much smaller than $\eta$.  In fact, they 
found that for separations $r \ll \eta$, the radial distribution function (RDF), an important 
measure of clustering, followed a power law given by
\begin{equation}
g(r) = c_0 \left ( \frac{\eta}{r} \right )^{c_1}
\end{equation}
where $g(r)$ is the RDF. 
The existence of power law for $r/\eta \approx 10^{-3}$ in the DNS 
of \citet{reade2000effect} suggests that the mechanism  
of preferential concentration alone is insufficient to explain clustering at 
such small separations.  

\citet{chun2005} investigated the continued clustering of monodisperse particles at 
sub-Kolmogorov separations, wherein we developed 
a theory for the RDF of low $St_\eta$, non-settling ($Fr \rightarrow \infty$) 
particle pairs.  
Motivated by the observation that much of the growth of the RDF occurs for 
separations $r < \eta$, \citet{chun2005} focused on the dynamics of pair separations 
in the disspation regime of turbulence.  Analytical closures were derived for the drift and 
diffusion fluxes in the PDF equation of pair relative positions.  The balance of these two fluxes determines the 
steady state value of the RDF at a given separation.  Of particular interest 
in that theory is the closure form for the drift flux $q_i^d(r)$ of 
monodisperse pairs, given by 
\begin{equation}
\label{eq:chun_drift}
q_i^d(r) = -\frac{St_\eta^2}{3} r_i ~\langle P \rangle (\bs{r})~
\int_{-\infty}^{t} \left \langle [S^2(t) - R^2(t)]~[S^2(t') - R^2(t')] \right \rangle~dt'
\end{equation}
where $S^2 = S_{ij} S_{ij}$ and $R^2 = R_{ij} R_{ij}$ are the second invariants 
of the strain-rate and rotation-rate tensors, respectively, along particle paths. 

It is evident from \eqref{eq:chun_drift} that the net drift flux will be negative 
or radially inward provided the primary particles sample more strain than 
rotation along their trajectories, a mechanism referred to as preferential concentration.   
One can also deduce from \eqref{eq:chun_drift} a second mechanism 
of clustering that is particularly relevant for sub-Kolmogorov scale separations.  
We can see from \eqref{eq:chun_drift} that the drift flux will continue to be 
negative even for $r < \eta$ provided we have a positive two-time correlation 
of $[S^2(t) - R^2(t)]$ along the trajectory of the primary particle.  
Thus, the sub-Kolmogorov scale clustering is driven by a path-history effect 
in that the pair separation at time $t$ continues to be influenced by the 
preferential sampling of strain-rate over rotation-rate by the primary particle 
at earlier times (and at larger separations, on average).  It is this path history 
effect that is responsible for the power-law behavior of the RDF at $r \ll \eta$. 
To the authors' knowledge, the \citet{chun2005} study   
is the first to provide an explicit relation for this effect through the 
integral in \eqref{eq:chun_drift}.

\citet{chun2005} also derived the drift-diffusion equation of the radial distribution 
function (RDF) for bidisperse, non-settling pairs.  Bidispersity, or more generally 
polydispersity, of the particle population is a key factor in determining 
clustering, and thereby the rate of particle collisions.  Bidispersity is also important when 
considering the effects of gravitational settling, since differential sedimentation 
is thought to be a key contributing factor to enhanced collision frequency.  In 
the current study, we consider a monodisperse population of settling particles.  
However,  our theory accounts for the effects of gravity through the modified sampling 
of turbulence by the settling particles.  
Although cloud droplets would be polydisperse, it is noteworthy that: (a) condensation tends to narrow the size distribution; (b) turbulence-induced coalescence is most important for nearly equal-sized drops 
for which differential sedimentation is weak, and (c) clustering is strongest for nearly equal-sized drops.  
In the rapid settling limit, particles 
experience an essentially frozen turbulence, so that the flow  
time scales along particle trajectories may be approximated as the Eulerian correlation 
length scales divided by the particle terminal velocity.

A detailed review of stochastic theories for the relative motion of inertial 
particle pairs is provided in \citet{rani2014} and \citet{rani2017}.  An important  
study is that of \citet{zaichik03}, who developed a stochastic theory for 
describing the relative velocities and positions of monodisperse particle pairs.  Their  
theory was conceived to be applicable for all Stokes numbers and for  
pair separations spanning all three regimes of turbulence, i.e.,~the integral, inertial 
and dissipation scale ranges.  Zaichik \& Alipchenkov derived a closure for the 
phase space diffusion current by using the 
Furutsu-Novikov-Donsker (FND) formula.  The FND formula relates 
the diffusion current to a series expansion in the 
cumulants of the fluid relative velocities 
seen by the pairs ($\Delta \boldsymbol{u}$) multiplied by 
the functional derivatives of the PDF with respect to 
$\Delta \boldsymbol{u}$ \citep{bragg2014a}.
%They further assumed that $\Delta \boldsymbol{u}$ had a Gaussian distribution, for 
%which the series expansion exactly reduces to only the second-order cumulant 
%of $\Delta \boldsymbol{u}$ multiplied by the first functional derivative 
%of the PDF \citep{bragg2014a}.     
\citet{zaichik03} then computed the statistics of pair separation and 
relative velocity by solving the equations for the zeroeth, first and 
second relative-velocity moments of the master PDF equation. 

\citet{bragg2014a} performed a rigorous, quantitative 
comparison of the \citet{chun2005} and \citet{zaichik2007} stochastic models for 
inertial pair dynamics in isotropic turbulence.  The focus of the Bragg \& Collins study  
was to compare and analyze the predictions of particle clustering at sub-Kolmogorov scale 
separations by the two theories.  The \citet{zaichik2007} study 
improved upon their earlier study \citep{zaichik03} by accounting 
for the unequal Lagrangian correlation timescales of the strain-rate and 
rotation-rate tensors.  Bragg \& Collins showed 
that the power-law exponents in the RDFs predicted by the two theories were in good agreement 
for $St_\eta \ll 1$ at $r \ll \eta$.  Through a detailed theoretical analysis, they
proved that this agreement was a consequence of the 
Chun {\it et al.}~drift velocity being the same as the 
leading order term in the \citet{zaichik2007} drift velocity.  
%This drift 
%velocity is given as
%\begin{eqnarray}
%\bs{v}^{\rm drift} = -\frac{1}{3} St_{\eta} \tau_{\eta} \left \langle
%S_p^2 - R_p^2
%\right \rangle
%\end{eqnarray}
%where $S_p^2$ and $R_p^2$ are the second invariants of strain-rate and rotation-rate  
%along inertial particle trajectories.  
As is to be expected, for $St_\eta \sim 1$, the theories diverge.  

In a recent analytical study, \citet{fouxon2015} considered the clustering behavior 
of fast-sedimenting particles in isotropic turbulence.  For a broad range of Stokes 
numbers ($St_\eta \gtrsim 1$, $St_\eta \ll 1$) and small Froude numbers ($Fr \ll 1$), they 
derived the power-law exponents characterizing the dependence of pair clustering on 
separation $r$.  The exponent that is applicable in the same parametric regime as  
in our study is \citep{fouxon2015}
\begin{eqnarray}
D_{KY} = \frac{4 \tau_\eta^2 \int_{0}^{\infty} \kappa^3 E_p(\kappa)~d\kappa} 
{\int_{0}^{\infty} \kappa E(\kappa)~d\kappa} \propto St_\eta^2
\end{eqnarray}
where $D_{KY}$ is the Lyapunov power-law exponent (known as the 
Kaplan-Yorke codimension), $E(\kappa)$ is the energy spectrum of isotropic 
turbulence, and $E_p(\kappa)$ is the spectrum of pressure fluctuations.  It may be noted 
that $D_{KY}$ scales as $St_\eta^2$, and is independent of $Fr$.  The exponent $\beta$ 
derived in the current study also shows the same dependence on $St_\eta$.  In our 
study, the first drift closure results in a $\beta$ that is independent of 
$Fr$.  However, the second drift closure can include the effects of $Fr$ through 
the two-time correlations of dissipation rate and enstrophy along particle 
trajectories.  \citet{fouxon2015} did not quantify $D_{KY}$, as 
the spectrum $E_p(\kappa)$ is not known.
In our study, however, $\beta_2$ is both quantified and compared with DNS data.

%A recent computational study of the dynamics of inertial particles 
%is that of \citet{peter2015b}.  They undertook a detailed study of the effects of 
%gravity on the dynamics of single particles, as well as particle 
%pairs, through direct numerical simulations 
%(DNS) of forced isotropic turbulence. They considered a wide range of Taylor 
%micro-scale Reynolds numbers ($88 \leq$ \Rel $\leq 597$), settling 
%velocity parameters ($0 \leq Sv_\eta \leq 100$) and 
%Stokes numbers ($0 \leq St_\eta \leq 56.2$).   
%For $St_\eta < 1$, they showed that the principal effect of gravity 
%on particle clustering is to decrease the inward (radial) drift, thereby 
%reducing the RDFs.
%They also found that gravity mitigates the preferential concentration mechanism by reducing  
%the interaction time between particles and the underlying turbulence.  
%Specifically, gravity reduces the Lagrangian time scales of strain-rate and 
%rotation-rate along the particle trajectories.  The effects of these time scales on 
%the inward drift flux is also evident in \eqref{eq:chun_drift}.   
%Anisotropy in particle clustering due to gravity was demonstrated and quantified through 
%the use of spherical harmonic functions to represent the RDF dependence on the polar 
%angle $\theta$.

In this Part I paper, we present the derivation of closures for the drift and diffusion  
fluxes in the probability density function (PDF) equation for pair separations $\bs{r}$ 
of rapidly setting, low-inertia, monodisperse particle pairs in isotropic turbulence.  
This study extends the \citet{chun2005} work by including the effects of particle 
settling in high gravity conditions.  Motivated by the \citet{chun2005} 
study, we approximate the fluid velocity field following the primary particle as 
locally linear.  An additional assumption regarding the fluid velocity gradient 
``seen" by the primary particle is also necessitated to resolve the third and fourth 
moments of the velocity gradient that appear in the drift flux.  Two types of assumption
regarding the velocity gradient lead to two separate closures for the drift flux, while 
the diffusion flux has only one closure.  The first closure of the drift flux entails 
assuming the ``seen" fluid velocity gradient to be Gaussian, while in the second, the 
scaled strain-rate and rotation-rate tensors ``seen" by the primary particle are assumed 
to be normally distributed.  In addition to the closures, an analytical solution is 
also derived for the PDF $\langle P \rangle (r,\theta)$, allowing us to quantify both the $r$-dependence 
and the anisotropy of clustering due to gravity.

The organization of the paper is as follows. Section \ref{sec:theory} presents 
the stochastic theory, including the derivation of the drift and diffusion flux closures. 
In section \ref{sec:solution}, analytical solution to the PDF 
$\langle P \rangle (r,\theta)$ is derived, with a power law dependence on 
$r$.  The results 
obtained from the first drift closure (in conjunction with the diffusion closure) are presented 
in section \ref{sec:results}.  
These results are based on using the analytical form of the energy 
spectrum that is valid in the high Reynolds-number limit.  The advantages 
of using this spectrum are that it obviates the need for DNS inputs, and 
importantly allows us to quantify the drift and diffusion fluxes in a 
universal manner (i.e., independent of $Re_\lambda$).  
Section \ref{sec:conclusions} summarizes the key findings of the Part I paper. 

\section{Stochastic Theory}
\label{sec:theory}
\nid In this section, we derive closure approximations for 
the drift and diffusion fluxes in the PDF equation for the relative positions  
$\bs{r}$ of monodisperse, low-inertia particle pairs that are settling rapidly 
in stationary isotropic turbulence. 
The theory is applicable in the $Fr \ll St_\eta \ll 1$ regime, and for  
pair separations in the dissipation regime of turbulence, i.e.,~$r < \eta$, where 
$\eta$ is the Kolmogorov length scale.  This restriction, however, allows us to approximate 
the fluid velocity field as being locally linear.  The effects of hydrodynamic and interparticle interactions on 
pair probability are neglected.  

We begin with the drift-diffusion equation derived by Chun {\it et al.} (2005) for the 
PDF $\langle P\rangle({\boldsymbol{r}};t) $:
\begin{eqnarray}
\frac{\partial \langle P\rangle }{\partial t}
+ \frac{\partial}{\partial r_i} \left (q_i^d + q_i^D \right ) = 0 \label{Eqn:PDF}
\end{eqnarray}
where the drift flux
\begin{equation}\label{eq:drift_flux}
q_i^d(\boldsymbol{r},t) = -\int_{-\infty}^{t} \left \langle W_i (\boldsymbol{r},\boldsymbol{x};t) ~
\frac{\partial W_l}{\partial r_l} [\boldsymbol{r}(t'),\boldsymbol{x}(t');t'] \right \rangle 
\langle P\rangle(\boldsymbol{r}';t') ~ dt',
\end{equation}
and the diffusive flux 
\begin{equation}\label{eq:diff_flux}
q_i^D(\boldsymbol{r},t) = -\int_{-\infty}^{t} \left \langle W_i (\boldsymbol{r},\boldsymbol{x};t) ~
W_j [\boldsymbol{r}(t'),\boldsymbol{x}(t');t'] \right \rangle 
\frac{\partial \langle P\rangle}{\partial r_j'}(\boldsymbol{r}';t') ~ dt'.
\end{equation}
In \eqref{eq:drift_flux} and \eqref{eq:diff_flux}, $\boldsymbol{r}' = \boldsymbol{r}(t')$ 
is the pair separation at time $t'$, and 
$\boldsymbol{x} = \boldsymbol{x}(t)$ is the primary particle 
position at time $t$. As the drift and diffusion fluxes at $\boldsymbol{r}$
depend on the pair probability and its derivative, respectively, at earlier pair separations 
$\boldsymbol{r}'$, equation (\ref{Eqn:PDF}) is non-local and accounts for the
path history effects.

The governing equations for the relative position (separation vector) $r_i$ and 
relative velocity $W_i$ of a settling, like-particle pair are:
\begin{equation}
\label{eq:drdt}
\frac{dr_i } {dt } = W_i
\end{equation}
\begin{align}\label{eq:dwdt}
\frac{dW_i} {dt } &= -\frac{1}{\tau_v} \left [ W_i(t) - 
\Delta u_i(\boldsymbol{r}(t),\boldsymbol{x}(t);t) \right ] \\
&\approx  
-\frac{1}{\tau_v} \left [ W_i(t) - 
\Gamma_{ik}(\boldsymbol{x}(t);t)~ r_k \right ]
\end{align}
where $\boldsymbol{x}(t)$ is the location of the primary particle, and 
$\Delta u_i(\boldsymbol{r}(t),\boldsymbol{x}(t);t)$ is the difference 
in the fluid velocities seen by the secondary and primary particles of a 
pair.  Using the approximation of a locally linear flow field, we write 
$\Delta u_i \approx \Gamma_{ik} r_k$, where $\Gamma_{ik} = \partial u_i/\partial x_k$ 
is the fluid velocity gradient at the location of the primary particle, $\boldsymbol{x}(t)$. 
In the case of monodisperse particle pairs, gravity influences 
pair relative motion only through the modified sampling of    
fluid velocity gradient by the primary particle.  

%In the rapidly settling regime, i.e., when $Sv_\eta = (g\tau_v)/u_\eta \gg 1$, the change 
%in time of the primary-particle position $\bs{x}$ may be approximated as 
%$\boldsymbol{x}(t') \approx \boldsymbol{x}(t) + \boldsymbol{g} \tau_v (t'-t)$, where 
%$\boldsymbol{g}$ is the gravity vector.  Furthermore, for $(t - t') < \tau_v$, we 
%may write $\bs{r}(t') \approx \bs{r}(t) = \bs{r}$.  Accordingly, equations 
%(\ref{eq:drift_flux}) and (\ref{eq:diff_flux}) for the drift and diffusion fluxes may be rewritten as
%\begin{equation}\label{eq:drift_flux2}
%q_i^d(\boldsymbol{r},t) = -\int_{-\infty}^{t} \left \langle W_i [\boldsymbol{r},\boldsymbol{x},t] ~
%\frac{\partial W_l}{\partial r_l} [\boldsymbol{r},\boldsymbol{x} + \boldsymbol{g} \tau_v (t'-t),t] \right \rangle 
%\langle P\rangle(\boldsymbol{r};t) ~ dt'
%\end{equation}
%and 
%\begin{equation}\label{eq:diff_flux2}
%q_i^D(\boldsymbol{r},t) = -\int_{-\infty}^{t} \left \langle W_i [\boldsymbol{r},\boldsymbol{x},t] ~
%W_j [\boldsymbol{r},\boldsymbol{x} + \boldsymbol{g} \tau_v (t'-t),t] \right \rangle 
%\frac{\partial \langle P\rangle}{\partial r_j}(\boldsymbol{r};t) ~ dt'
%\end{equation}
We now discuss the modeling of the drift and diffusion fluxes.  Two separate 
closures will be considered for the drift flux, whereas a single closure is obtained for the 
diffusion flux.  The two drift closures, DF1 and DF2, differ in the nature of the approximation 
made to analytically resolve the moments of the fluid velocity gradient tensor.  It will be seen 
that DF2 has the advantage of capturing key mechanisms of particle clustering.

\subsection{Drift Flux Closure 1 (DF1)}
Based on \citet{chun2005}, we express the pair relative velocity $W_i$ as a perturbation 
expansion with the Stokes number $St_\eta$ as the small parameter, as follows.
\begin{equation}
W_i = W_i^{[0]} + St_\eta W_i^{[1]} + \ldots
\end{equation}
Substituting this expansion into \eqref{eq:dwdt} and equating terms of equal 
order in $St_\eta$ yields
\begin{align}
& W_i^{[0]} = \Gamma_{ik} r_k \\
& W_i^{[1]} = -\frac{1}{\Gamma_\eta} \left [\frac{d \Gamma_{ik}}{dt}  + \Gamma_{ij} \Gamma_{jk} 
\right ]r_k
\end{align}
where $\Gamma_\eta = 1/\tau_\eta$ is the inverse of the Kolmogorov time scale $\tau_\eta$.  We have also  
used $dr_k/dt \approx W_k^{[0]}$ in deriving the expression for $W_i^{[1]}$. 
Thus, we can write
\begin{equation} \label{eq:w_i}
W_i (\boldsymbol{r}(t),\boldsymbol{x}(t);t) = \Gamma_{ik}(\boldsymbol{x}(t);t)~ r_k - 
\frac{St_\eta}{\Gamma_\eta}~ \left [\frac{d\Gamma_{ik}}{dt} + 
\Gamma_{ij}(\boldsymbol{x}(t),t)~\Gamma_{jk}(\boldsymbol{x}(t);t)
\right ] r_k
\end{equation}
\begin{equation} \label{eq:dwl_drl}
\frac{\partial W_l} {\partial r_l}(\boldsymbol{r}(t'),\boldsymbol{x}(t');t') = 
\Gamma_{ll} - \frac{St_\eta}{\Gamma_\eta}~ \left [ 
\frac{d\Gamma_{ll}}{dt} + \Gamma_{lm} \Gamma_{ml}
\right ] = - \frac{St_\eta}{\Gamma_\eta}~ 
\Gamma_{lm}(\boldsymbol{x}(t');t')~\Gamma_{ml}(\boldsymbol{x}(t');t')
\end{equation}
where $\Gamma_{ll} = 0$ due to continuity.   

Since the Stokes numbers of interest are small ($St_\eta \ll 1$), 
the fluid velocity gradients seen by the primary particle will be 
replaced by those of a collocated fluid particle.  With this approximation, we  
substitute \eqref{eq:w_i} and \eqref{eq:dwl_drl} 
into the drift flux given by \eqref{eq:drift_flux}, yielding 
\begin{gather}
q_i^d(\boldsymbol{r},t) = -\langle P\rangle(\boldsymbol{r};t)~ r_k 
\int_{-\infty}^{t} \Biggl \{ -\frac{St_\eta} {\Gamma_\eta} 
\left \langle \Gamma_{ik}(t) ~
\Gamma_{lm}(t') ~ \Gamma_{ml}(t') \right \rangle + \nonumber \\
\frac{St_\eta^2} {\Gamma_\eta^2} \left [
 \langle \frac{d\Gamma_{ik}}{dt}(t) ~ 
\Gamma_{lm}(t') ~ \Gamma_{ml}(t') \rangle + 
\left \langle \Gamma_{ij}(t) ~ \Gamma_{jk}(t) ~
\Gamma_{lm}(t') ~ \Gamma_{ml}(t') \right \rangle 
\right ]
\Biggr \} 
~ dt' \label{eq:drift_flux_gamma_terms_first}
\end{gather}
where $\Gamma_{ij}(t)$ and $\Gamma_{ij}(t')$ are the fluid velocity gradients at $t$ and $t'$ seen  
by a fluid particle at the same location as the inertial particle.  In 
\eqref{eq:drift_flux_gamma_terms_first}, $r_k$ and $\langle P\rangle(\boldsymbol{r};t)$  
have been brought out of the integral. This is reasonable given 
the parametric limits under consideration, and can be explained as follows. 
In the rapid settling limit, the correlation times of $\Gamma_{ij}$ along 
particle trajectories scale as $\eta/g\tau_v$, whereas pair separation 
evolves over $\tau_v \gg \eta/g\tau_v$.  Thus, the pair separation remains 
essentially unchanged during the time the velocity gradient remains 
correlated.  This allows us to pull $r_k$ out of the ensemble 
averaging $\langle \cdots \rangle$, as well as the time integral.  
Further, we are able to write 
$\langle P\rangle(\boldsymbol{r}';t') \approx \langle P\rangle(\boldsymbol{r};t)$, and then bring the PDF  
out of the time integral.  In Section \eqref{subsec:result_pdf}, we will explicitly quantify the times over 
which the PDF $\langle P\rangle$ evolves, and show that this is $\gg \eta/g\tau_v$, implying  
that the PDF is relatively unchanged during the $\Gamma_{ij}$ correlation times.

The drift flux in \eqref{eq:drift_flux_gamma_terms_first} contains the time integral of the third and 
fourth moments of fluid velocity gradient tensor along fluid particle 
trajectories.  To analytically resolve these moments, we apply the approximation 
that the velocity gradient tensor $\boldsymbol{\Gamma}$ is Gaussian.  {\it The resulting 
closure is referred to as DF1.}
%\citet{zaichik03} too invoked 
%an analogous approximation, wherein they assumed the fluid velocity differences 
%between the secondary and primary particles, $\bs{\Delta u}$ to be Gaussian.  While 
%\citet{zaichik03} recognized that $\bs{\Delta u}$ does not follow a Gaussian 
%distribution, they argued that the wide tails of the $\bs{\Delta u}$ distribution 
%should not significantly influence the lower order moments of the pair PDF such as the radial 
%distribution function.
Consequently, the two triple 
moment terms on the RHS of \eqref{eq:drift_flux_gamma_terms} would drop out.  Further, the fourth moment 
term may be written in terms of second moments as follows:
\begin{eqnarray} \label{eq:gamma_expn_gaussian}
\left \langle \Gamma_{ij}(t) ~ \Gamma_{jk}(t) ~
\Gamma_{lm}(t') ~ \Gamma_{ml}(t') \right \rangle &=& \nonumber 
\left \langle \Gamma_{ij}(t) ~ \Gamma_{jk}(t) \right \rangle 
\left \langle \Gamma_{lm}(t') ~ \Gamma_{ml}(t') \right \rangle  \nonumber \\
& + & 2\left \langle \Gamma_{ij}(t) ~ \Gamma_{lm}(t') \right \rangle 
\left \langle \Gamma_{jk}(t) ~ \Gamma_{ml}(t') \right \rangle 
\end{eqnarray}
The first term on the RHS of \eqref{eq:gamma_expn_gaussian} can be 
resolved by writing $\Gamma_{lm} = S_{lm} + R_{lm}$, where 
$S_{lm}$ and $R_{lm}$ are the fluid strain-rate and rotation-rate tensors.  Thus, we have
\begin{equation}
\left \langle \Gamma_{ij}(t) ~ \Gamma_{jk}(t) \right \rangle 
\left \langle \Gamma_{lm}(t') ~ \Gamma_{ml}(t') \right \rangle =
\left \langle \Gamma_{ij}(t) ~ \Gamma_{jk}(t) \right \rangle   
\left \langle S^2(t') - R^2(t') \right \rangle = 0
\label{eq:s2r2eqzero}
\end{equation}
since $\langle S^2 - R^2 \rangle = 0$ for fluid particles, 
where $S^2 = S_{lm} S_{lm}$ and $R^2 = R_{lm} R_{lm}$. 

Let us now consider the second term on the RHS of \eqref{eq:gamma_expn_gaussian}: 
\begin{equation}
2 \underbrace{\left \langle \Gamma_{ij}(t) ~ 
\Gamma_{lm}(t') \right \rangle }_{\circled{\rm {I}}} ~
\underbrace{\left \langle \Gamma_{jk}(t) ~ 
\Gamma_{ml}(t') \right \rangle }_{\circled{\rm {II}}}
\end{equation}
We will analyze the correlations \circled{\rm {I}} and \circled{\rm {II}} separately.  
In the rapid settling limit, particles fall through Kolmogorov-scale eddies 
in the time $\eta/(g\tau_v) \ll \tau_\eta$.  This enables us to express the two-time 
correlation of fluid velocity gradients as a two-point correlation with a spatial 
separation of $\boldsymbol{x}_g = \boldsymbol{g} \tau_v (t'-t)$.  Therefore, 
\begin{eqnarray}
\circled{\rm {I}} = \left \langle \Gamma_{ij}(\boldsymbol{x}(t),t) ~  \Gamma_{lm}(\boldsymbol{x}(t'),t') \right \rangle 
&=& \left \langle \Gamma_{ij}(\boldsymbol{x}(t),t) ~  \Gamma_{lm}(\boldsymbol{x}(t) + \boldsymbol{x}_g ,t) 
\right  \rangle \notag \\
&=& \left \langle \frac{\partial u_i} {\partial x_j} (\boldsymbol{x},t) ~
\frac{\partial u_l} {\partial x_m} (\boldsymbol{x} + \boldsymbol{x}_g,t) 
\right \rangle .
\end{eqnarray}

Expressing fluid velocities $u_i$ and $u_l$ in terms of Fourier coefficients 
in the wavenumber space yields
\begin{align}
\frac{\partial u_i} {\partial x_j} (\boldsymbol{x}) = 
\int \mi \kappa_j \widehat{u}_k (\boldsymbol{\kappa}) ~ e^{\mi \boldsymbol{\kappa} \bcdot \boldsymbol{x}}~
d\boldsymbol{\kappa} \\
\frac{\partial u_l} {\partial x_m} (\boldsymbol{x} + \boldsymbol{x}_g) = 
\int \mi \kappa'_m \widehat{u}_l (\boldsymbol{\kappa}') ~ e^{\mi \boldsymbol{\kappa}' \bcdot (\boldsymbol{x} + \boldsymbol{x}_g)}~
d\boldsymbol{\kappa}'
\end{align}
where $\mi = \sqrt{-1}$.  

Using the spatial homogeneity of fluid particle statistics, we can further average 
the correlation in \circled{\rm {I}} over $\boldsymbol{x}$-space giving \citep{pope_2000}
\begin{eqnarray}
%\left \langle \frac{\partial u_i} {\partial x_j} (\boldsymbol{x}) ~
%\frac{\partial u_l} {\partial x_m} (\boldsymbol{x} + \boldsymbol{x}_g) 
%\right \rangle &=& 
\circled{\rm {I}} = 
\left \langle \left \langle \frac{\partial u_i} {\partial x_j} (\boldsymbol{x}) ~
\frac{\partial u_l} {\partial x_m} (\boldsymbol{x} + \boldsymbol{x}_g) 
\right \rangle_{\mathcal{L}} \right \rangle 
&=& -\int \int d\boldsymbol{\kappa}~d\boldsymbol{\kappa}' ~\kappa_j \kappa'_m ~
\langle \widehat{u}_i (\boldsymbol{\kappa}) \widehat{u}_l (\boldsymbol{\kappa}') \rangle ~ 
\left \langle e^{\mi \boldsymbol{\kappa} \bcdot \boldsymbol{x}}
e^{\mi \boldsymbol{\kappa}' \bcdot (\boldsymbol{x} + \boldsymbol{x}_g)} \right \rangle_{\mathcal{L}} \nonumber \\
&=& -\int \int d\boldsymbol{\kappa}~d\boldsymbol{\kappa}' ~\kappa_j \kappa'_m ~
\langle \widehat{u}_i (\boldsymbol{\kappa}) \widehat{u}_l (\boldsymbol{\kappa}') \rangle ~ 
\delta(\boldsymbol{\kappa} + \boldsymbol{\kappa}')~ e^{\mi \boldsymbol{\kappa}' \bcdot \boldsymbol{x}_g}\nonumber \\
%&=& \int d\boldsymbol{\kappa} ~\kappa_j \kappa_m ~
%\langle \widehat{u}_i (\boldsymbol{\kappa}) \widehat{u}_l (\boldsymbol{-\kappa}) \rangle ~ 
%e^{-\mi \boldsymbol{\kappa} \bcdot \boldsymbol{x}_g} \nonumber \\
&=& \int d\boldsymbol{\kappa} ~\kappa_j \kappa_m ~
\langle \widehat{u}_i (\boldsymbol{\kappa}) \widehat{u}_l^* (\boldsymbol{\kappa}) \rangle ~ 
e^{-\mi \boldsymbol{\kappa} \bcdot \boldsymbol{x}_g} \nonumber \\
&=& \int d\boldsymbol{\kappa} ~\kappa_j \kappa_m ~
\Phi_{il}(\boldsymbol{\kappa}) ~ 
e^{-\mi \boldsymbol{\kappa} \bcdot \boldsymbol{x}_g} \label{eq:corr_term_I}
\end{eqnarray}
where $\langle ... \rangle_{\mathcal{L}}$ denotes averaging over $\bs{x}$, 
$\delta(\cdots)$ denotes the Diract delta function, 
$\boldsymbol{\kappa}$ and $\boldsymbol{\kappa}'$ are both wavenumber vectors, 
$\widehat{u}_i (\boldsymbol{\kappa})$ is a Fourier component of the fluid 
velocity corresponding to the wavenumber $\boldsymbol{\kappa}$, and 
$\widehat{u}_l^*$ is the complex conjugate of $\widehat{u}_l$.  The velocity 
spectrum tensor $\Phi_{il}(\boldsymbol{\kappa})$ can be written in terms 
of energy spectrum $E(\kappa)$ \citep{pope_2000}
\begin{eqnarray}
\Phi_{il}(\boldsymbol{\kappa}) &=& \frac{E(\kappa)}{4 \pi \kappa^2} 
\left (\delta_{il} - \frac{\kappa_i \kappa_l}{\kappa^2} \right )
\label{eq:velspec}
\end{eqnarray}
Similarly, 
\begin{eqnarray}
\circled{\rm {II}} = 
\left \langle \Gamma_{jk}(\boldsymbol{x}(t),t) ~  \Gamma_{ml}(\boldsymbol{x}(t) + \boldsymbol{x}_g ,t) \right \rangle &=& 
\left \langle \frac{\partial u_j} {\partial x_k} (\boldsymbol{x}) ~
\frac{\partial u_m} {\partial x_l} (\boldsymbol{x} + \boldsymbol{x}_g) 
\right \rangle \nonumber \\
&=& \int d\boldsymbol{\kappa}' ~\kappa'_k \kappa'_l ~
\Phi_{jm}(\boldsymbol{\kappa'}) ~ 
e^{-\mi \boldsymbol{\kappa'} \bcdot \boldsymbol{x}_g}
\end{eqnarray}

The time integral of the product of \circled{\rm {I}} and \circled{\rm {II}} is
\begin{eqnarray} 
\int_{-\infty}^{0} dt \left \langle \Gamma_{ij}(\boldsymbol{x}(0),0) ~  \Gamma_{lm}(\boldsymbol{x}(0) + \boldsymbol{x}_g ,0) \right \rangle 
\left \langle \Gamma_{jk}(\boldsymbol{x}(0),0) ~  \Gamma_{ml}(\boldsymbol{x}(0) + \boldsymbol{x}_g ,0) 
\right \rangle \nonumber \\
= \int \int d\boldsymbol{\kappa} d\boldsymbol{\kappa}' ~\kappa_j \kappa_m ~
\Phi_{il}(\boldsymbol{\kappa}) ~ \kappa'_k \kappa'_l ~
\Phi_{jm}(\boldsymbol{\kappa'}) \int_{-\infty}^{0} dt~ 
e^{-\mi (\boldsymbol{\kappa}+\boldsymbol{\kappa}') \bcdot \boldsymbol{g} \tau_v t} \nonumber \\
= \int \int d\boldsymbol{\kappa} d\boldsymbol{\kappa}' ~\kappa_j \kappa_m ~
\Phi_{il}(\boldsymbol{\kappa}) ~ \kappa'_k \kappa'_l ~
\Phi_{jm}(\boldsymbol{\kappa'}) 
\Biggl \{ \frac{1}{2} \delta \left [- \frac 
{(\boldsymbol{\kappa}+\boldsymbol{\kappa}') \bcdot \boldsymbol{g} \tau_v } 
{2\pi} \right ] -  
\frac{1}{\mi (\boldsymbol{\kappa}+\boldsymbol{\kappa}') \bcdot \boldsymbol{g} \tau_v}
\Biggr \}  \notag \\ \label{eq:product_I_II}
\end{eqnarray}
where we have used the Fourier transform identity for the 
time integral $\int_{-\infty}^{0} dt~ 
e^{-\mi (\boldsymbol{\kappa}+\boldsymbol{\kappa}') \bcdot \boldsymbol{g} \tau_v t}$.
Let us consider the two terms in the above integral separately.  The first term 
given by the integral 
\begin{eqnarray}\label{eq:int_term1}
\frac{1}{2} \int \int d\boldsymbol{\kappa} d\boldsymbol{\kappa}' ~\kappa_j \kappa_m ~
\Phi_{il}(\boldsymbol{\kappa}) ~ \kappa'_k \kappa'_l ~
\Phi_{jm}(\boldsymbol{\kappa'}) ~
\delta \left [- \frac 
{(\boldsymbol{\kappa}+\boldsymbol{\kappa}') \bcdot \boldsymbol{g} \tau_v } 
{2\pi} \right ]
\end{eqnarray}
is non-zero only when 
$(\boldsymbol{\kappa}+\boldsymbol{\kappa}') \bcdot \boldsymbol{g} = 0$, or 
$(\boldsymbol{\kappa}+\boldsymbol{\kappa}')$ is 
$\perp$ to $\boldsymbol{g} = -g \hat{\boldsymbol{e}}_3$.  Let 
$(\boldsymbol{\kappa}+\boldsymbol{\kappa}') = \boldsymbol{\xi} = (\xi_1,\xi_2,0)$  
such that this property is satisfied.   Using the sifting property of the Diract 
delta function, as well as the identity $\delta(ax) = (1/|a|) \delta(x)$, the integral 
in \eqref{eq:int_term1} now becomes 
\begin{gather}
\frac{1}{2} \frac{2 \pi}{g \tau_v } \int \int d\boldsymbol{\kappa}~ d\xi_1~ d\xi_2 ~\kappa_j \kappa_m ~
\Phi_{il}(\boldsymbol{\kappa}) ~ (\xi_k - \kappa_k) (\xi_l - \kappa_l) ~
\Phi_{jm}(\boldsymbol{\xi} - \boldsymbol{\kappa}) \label{eq:int_term1_eps}
\end{gather}
Next, we consider the second term in the integral in the last line of \eqref{eq:product_I_II}.  
Unlike the first term, it will be seen subsequently that this term does not make any contribution to the drift. 

Recognizing that the particles preferentially sample the 
velocity gradients along the $x_3$ or gravity direction, we apply 
the tensorial constraints for a field that is homogeneous along the 
$x_1$ and $x_2$ directions.  Expressing the integral in Eq.~\eqref{eq:int_term1_eps} 
in terms of these tensor constraints, we have
\begin{eqnarray}
\frac{ \pi}{g \tau_v } \int \int d\boldsymbol{\kappa}~ d\xi_1~ d\xi_2 ~\kappa_j \kappa_m ~
\Phi_{il}(\boldsymbol{\kappa}) ~ (\xi_k - \kappa_k) (\xi_l - \kappa_l) ~
\Phi_{jm}(\boldsymbol{\xi} - \boldsymbol{\kappa}) =  \notag \\
\lambda_1 \left (\delta_{ik} - \delta_{i3} \delta_{k3} \right ) + 
\lambda_2 \delta_{i3} \delta_{k3}
\end{eqnarray}
Multiplying the above equation with $(\delta_{ik} - \delta_{i3} \delta_{k3} )$ gives 
$\lambda_1$ and with $\delta_{i3} \delta_{k3}$ gives $\lambda_2$.
\begin{gather}
\lambda_1 = \frac{ \pi}{2g \tau_v } 
\int \int d\boldsymbol{\kappa}~ d\xi_1~ d\xi_2~ \kappa_j \kappa_m ~
\Phi_{jm}(\boldsymbol{\xi} - \boldsymbol{\kappa})~(\xi_l - \kappa_l) ~
\left [ \Phi_{il}(\boldsymbol{\kappa}) ~ (\xi_i - \kappa_i) + 
\Phi_{3l}(\boldsymbol{\kappa})~\kappa_3 \right ] \label{eq:lambda1} \\
\lambda_2 = -\frac{ \pi}{g \tau_v } 
\int \int d\boldsymbol{\kappa}~ d\xi_1~ d\xi_2~ \kappa_j \kappa_m ~
\Phi_{3l}(\boldsymbol{\kappa}) ~ \kappa_3 ~(\xi_l - \kappa_l) ~
\Phi_{jm}(\boldsymbol{\xi} - \boldsymbol{\kappa}) \label{eq:lambda2}
\end{gather}
Using spherical coordinates to represent the $\boldsymbol{\kappa}$ vector and 
cylindrical coordinates to represent $\boldsymbol{\xi}$, we have
\begin{gather} 
\boldsymbol{\kappa} = (\kappa_1, \kappa_2, \kappa_3) = 
(\kappa \sin \theta \cos \phi, \kappa \sin \theta \sin \phi, \kappa \cos \theta) \nonumber \\
\boldsymbol{\xi} = (\xi_1,\xi_2,0) = (\xi \cos \psi, \xi \sin \psi, 0) \label{eq:sphere_cylinder_coord}
\end{gather}
Using \eqref{eq:sphere_cylinder_coord} in the 
equations for $\lambda_1$ and $\lambda_2$, i.e. Eqs.~\eqref{eq:lambda1} and \eqref{eq:lambda2}, 
\begin{align}
\lambda_1 = \frac{\pi}{2g \tau_v} \int_{\phi=0}^{2\pi} d\phi \int_{\theta = 0}^{\pi} d\theta 
\int_{\kappa = 0}^{\infty} d\kappa \int_{\psi = 0}^{2\pi} d\psi \int_{\xi = 0}^{\infty} d\xi \times 
[\text{Integrand \circled{1}}] \label{eq:lamb1} \\
\lambda_2 = \frac{\pi}{g \tau_v} \int_{\phi=0}^{2\pi} d\phi \int_{\theta = 0}^{\pi} d\theta 
\int_{\kappa = 0}^{\infty} d\kappa \int_{\psi = 0}^{2\pi} d\psi \int_{\xi = 0}^{\infty} d\xi \times 
[\text{Integrand \circled{2}}] \label{eq:lamb2} 
\end{align}
where
\begin{gather}
\text{Integrand \circled{1}} = 
\frac{E(|\boldsymbol{\xi} - \boldsymbol{\kappa}|)}
{4 \pi \left [\xi^2 + \kappa^2 - 2\xi \kappa \sin \theta \cos(\psi - \phi) \right ]} ~
\frac{ \xi^3 \kappa^4 \left [1 - \sin^2 \theta \cos^2 (\psi - \phi) \right ] \sin \theta} 
{\xi^2 + \kappa^2 - 2\xi \kappa \sin \theta \cos(\psi - \phi)} \times \notag \\
 \frac{E(\kappa)}{4 \pi \kappa^2} 
\left \{ \xi^2 \left [1 - \sin^2 \theta \cos^2 (\psi - \phi) \right ] - 
\frac{\xi \kappa \cos \theta}{2} \sin2\theta \cos(\psi - \phi) \right \}
\end{gather}
\begin{gather}
\text{Integrand \circled{2}} = 
\frac{E(|\boldsymbol{\xi} - \boldsymbol{\kappa}|)}
{4 \pi \left [\xi^2 + \kappa^2 - 2\xi \kappa \sin \theta \cos(\psi - \phi) \right ]} ~
\frac{ \xi^3 \kappa^4 \left [1 - \sin^2 \theta \cos^2 (\psi - \phi) \right ] \sin \theta} 
{\xi^2 + \kappa^2 - 2\xi \kappa \sin \theta \cos(\psi - \phi)} \times \notag \\
\frac{E(\kappa)}{4 \pi \kappa^2} ~\frac{\xi \kappa \cos\theta}{2} ~\sin2\theta \cos(\psi - \phi)
 \label{eq:lambda2_final}
\end{gather}

Let us now consider the second term in the integral of Eq. \eqref{eq:product_I_II} 
(it has already been mentioned earlier that this term goes to zero), given by
\begin{align}
-\int \int d\boldsymbol{\kappa} d\boldsymbol{\kappa}' ~\kappa_j \kappa_m ~
\Phi_{il}(\boldsymbol{\kappa}) ~ \kappa'_k \kappa'_l ~
\Phi_{jm}(\boldsymbol{\kappa'}) 
\frac{1}{\mi (\boldsymbol{\kappa}+\boldsymbol{\kappa}') \bcdot \boldsymbol{g} \tau_v} = 
\lambda_3 \left (\delta_{ik} - \delta_{i3} \delta_{k3} \right ) + 
\lambda_4 \delta_{i3} \delta_{k3} \notag \\
= \lambda_3 \left (\delta_{ik} - \frac{g_i g_k}{g^2} \right ) + 
\lambda_4 \frac{g_i g_k}{g^2} 
\label{eq:2nd_integral_zero}
\end{align}
where $g_i$ is the gravity vector that is non-zero only when $i = 3$.  
The integral on the LHS of \eqref{eq:2nd_integral_zero} is odd in $\mathbf{g}$, but the 
RHS is even in $\mathbf{g}$.  Hence the integral will be zero.  
The final form of drift flux in DF1 is given by
\begin{eqnarray}\label{eq:drift_flux_gamma_terms_final}
q_i^d(\boldsymbol{r},t) = -\langle P\rangle(r,\theta)~ 2r_k ~
\frac{St_\eta^2}{\Gamma_\eta^2} ~[\lambda_1 \left (\delta_{ik} - \delta_{i3} \delta_{k3} \right ) + 
\lambda_2 \delta_{i3} \delta_{k3}]
\end{eqnarray} 
where $\theta$ is the spherical polar angle that accounts for the anisotropy in the radial 
distribution function (RDF), and $\lambda_1$ and $\lambda_2$ are given by 
\eqref{eq:lamb1} and \eqref{eq:lamb2}.

\subsection{Drift Flux Closure 2 (DF2)}
We now present the development of the second drift closure (DF2).  
It is evident from \eqref{eq:s2r2eqzero} that the first closure (DF1) does not capture the 
two-time autocorrelations and cross-correlations of the strain-rate and 
rotation-rate invariants---$\langle S^2(t) S^2(t') \rangle$, $\langle R^2(t) R^2(t') \rangle$, 
$\langle S^2(t) R^2(t') \rangle$ and $\langle R^2(t) S^2(t') \rangle$.   
As seen in \eqref{eq:chun_drift}, the drift flux of non-settling pairs  
involves the time integration of these correlations.  We anticipate that 
the mechanism(s) driving the accumulation of pairs for $Fr \ll 1$ 
will be related to those for $Fr \gg 1$ (zero gravity case), albeit modulated
by gravity.  Therefore, our objective 
is to derive a closure (DF2) that accounts for the above correlations.

The closures DF1 and DF2 differ in the assumption  
made to resolve the moments of the fluid velocity gradient tensor. 
In DF1, we had assumed  
the velocity gradient tensor to be Gaussian, whereas in DF2, we regard the 
dimensionless strain-rate and rotation-rate tensors to be normally distributed. 

Referring to the drift flux $q_i^d$ in \eqref{eq:drift_flux_gamma_terms_first}, we first  
decompose the velocity gradient tensor $\Gamma_{ij}(t)$ into the sum of the 
strain-rate and rotation-rate tensors, $S_{ij}(t)$ and $R_{ij}(t)$.  Subsequently, we 
non-dimensionalize $S_{ij}$ and $R_{ij}$ using the instantaneous dissipation rate and 
enstrophy, $\epsilon(t)$ and $\zeta(t)$ respectively.  These two steps allow us 
to write $\Gamma_{ij}(t)$ as
\begin{align}
\Gamma_{ij}(t) &= S_{ij}(t) + R_{ij}(t) \\
%&= \sqrt{ \frac{\epsilon(t)}{2\nu} } ~\frac{ S_{ij} }{\sqrt{ \frac{\epsilon(t)}{2\nu} }} + 
%\sqrt{ \frac{\zeta(t)}{2\nu} } ~\frac{ R_{ij} }{\sqrt{ \frac{\zeta(t)}{2\nu} }} 
%\label{eq:gamma_ij_inter} \\
&= \frac{1}{\sqrt{2\nu} } \left [
\sqrt{\epsilon(t)} ~\sigma_{ij}(t) + \sqrt{\zeta(t)} ~\rho_{ij}(t)
\right ] \label{eq:gamma_ij_norm}
\end{align}
where $\epsilon(t) = 2\nu S_{ij}(t) S_{ij}(t)$,  
$\zeta(t) = 2\nu R_{ij}(t) R_{ij}(t)$ [$\nu$ is the kinematic viscosity], and 
$\sigma_{ij}(t)$ and $\rho_{ij}(t)$ are the 
dimensionless strain-rate and rotation-rate tensors, respectively.
%\begin{align}
%\sigma_{ij}(t) = \frac{ S_{ij}(t) }{\sqrt{ \frac{\epsilon(t)}{2\nu} }} ~~~~~~~~
%\rho_{ij}(t) = \frac{ R_{ij}(t) }{\sqrt{ \frac{\zeta(t)}{2\nu} }}.
%\label{eq:strainrot_norm}
%\end{align}

Substituting \eqref{eq:gamma_ij_norm} for $\boldsymbol{\Gamma}$ 
in \eqref{eq:drift_flux_gamma_terms_first}, and assuming 
$\sigma_{ij}(t)$ and $\rho_{ij}(t)$ to be normally distributed, we can drop  
the third moments of $\boldsymbol{\Gamma}$ as they, in turn, give rise 
to third moments of $\boldsymbol{\sigma}$, $\boldsymbol{\rho}$, and to 
cross correlations of third order involving $\boldsymbol{\sigma}$ and $\boldsymbol{\rho}$.  With these 
simplifications, the drift flux in \eqref{eq:drift_flux_gamma_terms_first} reduces to  
\begin{gather}
q_i^d(\boldsymbol{r},t) = -\langle P\rangle(\boldsymbol{r};t)~\frac{St_\eta^2}{\Gamma_\eta^2}~ r_k 
\int_{-\infty}^{t} d_{ik}~dt' \label{eq:drift_flux_gamma_terms}
\end{gather}
where
\begin{gather}
d_{ik} = \left \langle \Gamma_{ij}(t) ~ \Gamma_{jk}(t) ~
\Gamma_{lm}(t') ~ \Gamma_{ml}(t') \right \rangle \approx \notag \\ \frac{1}{4\nu^2} \Bigl \{ ~
\left \langle \epsilon(t) ~\epsilon(t') \right \rangle [
\langle \sigma_{ij}(t) ~\sigma_{jk}(t) \rangle ~
\langle \sigma_{lm}(t') ~\sigma_{lm}(t') \rangle + %\notag \\
2\langle \sigma_{ij}(t) ~\sigma_{lm}(t') \rangle ~
\langle \sigma_{jk}(t) ~\sigma_{lm}(t') \rangle ] \notag \\
- \left \langle \epsilon(t) ~\zeta(t') \right \rangle
\langle \sigma_{ij}(t) ~\sigma_{jk}(t) \rangle ~
\langle \sigma_{lm}(t') ~\sigma_{lm}(t') \rangle 
+ \left \langle \zeta(t) ~\epsilon(t') \right \rangle
\langle \rho_{ij}(t) ~\rho_{jk}(t) \rangle ~
\langle \sigma_{lm}(t') ~\sigma_{lm}(t') \rangle \notag \\
-\left \langle \zeta(t) ~\zeta(t') \right \rangle [
\langle \rho_{ij}(t) ~\rho_{jk}(t) \rangle ~
\langle \rho_{lm}(t') ~\rho_{lm}(t') \rangle + 
2\langle \rho_{ij}(t) ~\rho_{lm}(t') \rangle ~
\langle \rho_{jk}(t) ~\rho_{lm}(t') \rangle ]
~\Bigr \}  \label{eq:gamma_fourthmom_sigmas}
\end{gather}
In \eqref{eq:gamma_fourthmom_sigmas}, we have also assumed that 
$\epsilon(t)$ and $\boldsymbol{\sigma}(t)$ are weakly correlated, and 
so are $\zeta(t)$ and $\boldsymbol{\rho}(t)$.  This is a reasonable 
approximation since the dissipation rate and enstrophy vary over 
characteristic time scales that are quite different from those 
of strain-rate and rotation-rate tensors, respectively.  The former two  
have scales of the order of large-eddy time scales \citep{chun2005}.  But, the components of  
strain rate  have time scales $\sim 2.3\tau_\eta$ and those of rotation rate 
$\sim 7.2\tau_\eta$ \citep{chun2005,zaichik2007}, where $\tau_\eta$ is the 
Kolmogorov time scale.  

Due to isotropy, the one-time correlations of the $\bs{\sigma}$ and $\bs{\rho}$ tensors 
in \eqref{eq:gamma_fourthmom_sigmas} can be written as \citep{chun2005}
\begin{gather}
%\langle \sigma_{ij}(t) ~\sigma_{jk}(t) \rangle = \frac{1}{10} 
%\left [\delta_{ij} \delta_{kl} + \delta_{il} \delta_{jk} - \frac{2}{3} \delta_{ik} \delta_{jl} \right ] \\
\langle \sigma_{ij}(t) ~\sigma_{jk}(t) \rangle = \frac{1}{3} \delta_{ik} \\
\langle \sigma_{lm}(t) ~\sigma_{lm}(t) \rangle = 1 \\
\langle \rho_{ij}(t) ~\rho_{jk}(t) \rangle = -\frac{1}{3} \delta_{ik} \\
\langle \rho_{lm}(t) ~\rho_{lm}(t) \rangle = 1
\end{gather}
We now have 
\begin{gather}
d_{ik} = \frac{1}{4\nu^2} \Bigl \{ ~\frac{1}{3} \delta_{ik}
\left [ \left \langle \epsilon(t) ~\epsilon(t') \right \rangle + 
\left \langle \epsilon(t) ~\zeta(t') \right \rangle - 
\left \langle \zeta(t) ~\epsilon(t') \right \rangle - 
\left \langle \zeta(t) ~\zeta(t') \right \rangle 
\right ] +  \notag \\
2 \left \langle \epsilon(t) ~\epsilon(t') \right \rangle \langle \sigma_{ij}(t) ~\sigma_{lm}(t') \rangle ~
\langle \sigma_{jk}(t) ~\sigma_{lm}(t') \rangle - \notag \\
2 \left \langle \zeta(t) ~\zeta(t') \right \rangle \langle \rho_{ij}(t) ~\rho_{lm}(t') \rangle ~
\langle \rho_{jk}(t) ~\rho_{lm}(t') \rangle
~\Bigr \}  \label{eq:dik2}
\end{gather}
In \eqref{eq:dik2}, we will express the two-time correlation of dissipation rate as \citep{chun2005}
\begin{equation}
\langle \epsilon(t) \epsilon(t') \rangle = \langle \epsilon^2 \rangle
\exp \left ( -\frac{t-t'}{T_{\epsilon \epsilon} } \right )
\label{eq:t_epseps}
\end{equation}
so that 
\begin{equation}
\int_{-\infty}^{t} \langle \epsilon(t) \epsilon(t') \rangle~dt' = \langle \epsilon^2 \rangle T_{\epsilon \epsilon}
\end{equation}
where $T_{\epsilon \epsilon}$ is the correlation time scale of $\epsilon$. 
In a similar manner, the correlations 
$\langle \epsilon(t) \zeta(t') \rangle$, $\langle \zeta(t) \epsilon(t') \rangle$ and 
$\langle \zeta(t) \zeta(t') \rangle$ are expressed in terms of the 
correlation time scales $T_{\epsilon \zeta}$, $T_{\zeta \epsilon}$ and 
$T_{\zeta \zeta}$, respectively.  Thus, we have
\begin{gather}
\int_{-\infty}^{t} d_{ik}~dt' = 
\frac{1}{4\nu^2} \Bigl \{ ~\frac{1}{3} \delta_{ik}
\left [ \langle \epsilon^2 \rangle T_{\epsilon \epsilon} + 
\langle \epsilon \zeta \rangle T_{\epsilon \zeta} - 
\langle \zeta \epsilon \rangle T_{\zeta \epsilon} - 
\langle \zeta^2 \rangle T_{\zeta \zeta} 
\right ] +  \notag \\
2 \langle \epsilon^2 \rangle \int_{-\infty}^{t} 
\exp \left ( -\frac{t-t'}{T_{\epsilon \epsilon} } \right ) \langle \sigma_{ij}(t) ~\sigma_{lm}(t') \rangle ~
\langle \sigma_{jk}(t) ~\sigma_{lm}(t') \rangle~dt' - \notag \\
2 \langle \zeta^2 \rangle \int_{-\infty}^{t} 
\exp \left ( -\frac{t-t'}{T_{\zeta \zeta} } \right ) \langle \rho_{ij}(t) ~\rho_{lm}(t') \rangle ~
\langle \rho_{jk}(t) ~\rho_{lm}(t') \rangle~dt'
~\Bigr \}
\label{eq:int_rhs_dik}
\end{gather}
%Strictly speaking, the time scales $T_{\epsilon \epsilon}$, $T_{\epsilon \zeta}$, 
%$T_{\zeta \epsilon}$ and $T_{\zeta \zeta}$ are defined along the 
%trajectories of inertial settling particles.  Thus, they allow us to retain in the theory the effects 
%of moderate $Fr$ on the particle clustering.
%These time scales may be obtained through DNS of settling particles, which is rather 
%expensive, particularly for small values of $Fr$.  Alternatively, 
In the rapid settling limit, the time scales $T_{\epsilon \epsilon}$, $T_{\epsilon \zeta}$, 
$T_{\zeta \epsilon}$ and $T_{\zeta \zeta}$ can be approximated as the ratio of the corresponding 
Eulerian correlation length and the particle terminal velocity.  For example,
\begin{equation}
T_{\epsilon \epsilon} \approx \frac{L_{\epsilon \epsilon}} {g\tau_v}
\end{equation}
where $L_{\epsilon \epsilon}$ is the Eulerian length scale of $\epsilon$.  The various Eulerian 
length scales are evaluated via DNS of isotropic turbulence.

To evaluate the two integrals on the RHS of \eqref{eq:int_rhs_dik}, we need to resolve 
the two-time correlations of $\bs{\sigma}$ and $\bs{\rho}$ --- $\langle \sigma_{ij}(t) ~\sigma_{lm}(t') \rangle$, 
$\langle \sigma_{jk}(t) ~\sigma_{lm}(t') \rangle$, 
$\langle \rho_{ij}(t) ~\rho_{lm}(t') \rangle$, and 
$\langle \rho_{jk}(t) ~\rho_{lm}(t') \rangle$.  
Analogous to the process leading to \eqref{eq:corr_term_I}, we will transform the two-time 
correlations of $\bs{\sigma}$ and $\bs{\rho}$ into two-point correlations 
with a spatial separation of $\bs{x}_g = \bs{g} \tau_v (t'-t)$, and  
express the two-point correlations as Fourier integrals.  
Subsequently, we apply the tensorial constraints 
arising from the particles sampling the flow field preferentially 
along the $x_3$ direction, but homogeneously in the $x_1-x_2$ plane. 
Accordingly, $\langle \sigma_{ij}(t) ~\sigma_{lm}(t') \rangle$ can be expressed as
\begin{gather}
\langle \sigma_{ij}(\bs{x},t) ~\sigma_{lm}(\bs{x}',t') \rangle  
= \int d\boldsymbol{\kappa} ~ \langle \widehat{\sigma}_{ij} (\bs{\kappa},t) ~
\widehat{\sigma}^{*}_{lm} (\bs{\kappa},t)\rangle ~ 
e^{-\mi \boldsymbol{\kappa} \bcdot \boldsymbol{x}_g} = \mathscr{L}_{ijlm} = \notag \\
\alpha_1 \delta_{ij} \delta_{lm} + \alpha_2 (\delta_{im} \delta_{jl} + \delta_{il} \delta_{jm}) + 
\alpha_4 \delta_{i3}\delta_{j3}\delta_{l3}\delta_{m3} + \notag \\
\alpha_5 (\delta_{i3}\delta_{j3} \delta_{lm} + \delta_{ij} \delta_{l3} \delta_{m3} ) + 
\alpha_6(\delta_{i3} \delta_{l3} \delta_{jm} + 
\delta_{i3} \delta_{m3} \delta_{jl} + \delta_{j3} \delta_{l3} \delta_{im} + 
\delta_{j3} \delta_{m3} \delta_{il} )
\label{eq:sig_ijlm}
\end{gather} 
where 
\begin{gather}
\alpha_1 = -\frac{1}{8} (2B_1 - B_2 - 4B_3); \hspace{0.1in} 
\alpha_2 = \frac{1}{8} (2B_1 + B_2 - 4B_3) \notag \\
\alpha_4 = \frac{1}{8} (2B_1 + 35 B_2 - 20 B_3); \hspace{0.1in} 
\alpha_5 = \frac{1}{8} (2B_1 - 5B_2 - 4B_3) \notag \\
\alpha_6 = -\frac{1}{8} (2B_1 + 5B_2 - 8B_3) \notag
\end{gather}
\begin{gather}
B_1 = \frac{\nu}{2 \langle \epsilon \rangle} 
\left [ \frac{1}{\pi} \int d\boldsymbol{\kappa} ~E(\kappa) 
~e^{-\mi \boldsymbol{\kappa} \bcdot \boldsymbol{x}_g} \right ] 
\label{eq:b1_text} \\
B_2 = \frac{\nu}{2 \langle \epsilon \rangle} 
\left [ 4 \int d\boldsymbol{\kappa} ~\kappa_3^2~\frac{E(\kappa) }{4\pi \kappa^2}
\left (1 - \frac{\kappa_3^2}{\kappa^2} \right )
e^{-\mi \boldsymbol{\kappa} \bcdot \boldsymbol{x}_g} \right ] \label{eq:b2_text} \\
B_3 = \frac{\nu}{2 \langle \epsilon \rangle} 
\left [ \int d\boldsymbol{\kappa} ~\kappa_j \kappa_j~\frac{E(\kappa) }{4\pi \kappa^2}
\left (1 + \frac{\kappa_3^2}{\kappa^2} \right )
e^{-\mi \boldsymbol{\kappa} \bcdot \boldsymbol{x}_g} \right ] \label{eq:b3_text} 
\end{gather}
In the equations \eqref{eq:b1_text}-\eqref{eq:b3_text}, $E(\kappa)$ is the 
energy spectrum of isotropic turbulence, and $\kappa_3$ is the 
component of $\boldsymbol{\kappa}$ along the $x_3$ direction.  Appendix \ref{sec:app_tensor} presents 
the process for determining the form of the tensorial constraints in \eqref{eq:sig_ijlm}, as well as 
the coefficients $\alpha_1, \alpha_2$ and others.
Appendix \ref{sec:app2} presents the evaluation of $\langle \widehat{\sigma}_{ij} (\bs{\kappa},t) ~
\widehat{\sigma}^{*}_{lm} (\bs{\kappa},t)\rangle$.

The term $\langle \sigma_{jk}(\bs{x},t) ~\sigma_{lm}(\bs{x}',t') \rangle$ 
may also be expressed analogous to \eqref{eq:sig_ijlm}. 
Thus, the product $\langle \sigma_{ij}(t) ~\sigma_{lm}(t') \rangle ~
\langle \sigma_{jk}(t) ~\sigma_{lm}(t') \rangle$ in \eqref{eq:int_rhs_dik} can now be written as 
\begin{align}
&\langle \sigma_{ij}(t) ~\sigma_{lm}(t') \rangle ~
\langle \sigma_{jk}(t) ~\sigma_{lm}(t') \rangle = 
\delta_{ik} (3 \alpha_1 \alpha_1 + 4 \alpha_1 \alpha_2 + 2 \alpha_1 \alpha_5 +  
8 \alpha_2 \alpha_2 + 4 \alpha_2 \alpha_6 + \notag \\ 
&\alpha_5  \alpha_5 + 2 \alpha_6 \alpha_6) + \delta_{i3} \delta_{k3} 
(2 \alpha_1 \alpha_4 + 6 \alpha_1 \alpha_5 + 8 \alpha_1 \alpha_6 +  
4 \alpha_2 \alpha_4 + 8 \alpha_2 \alpha_5 + \notag \\ 
&20\alpha_2  \alpha_6 + \alpha_4 \alpha_4 + 4 \alpha_4 \alpha_5 + 
8 \alpha_4 \alpha_6 + 5 \alpha_5 \alpha_5 + 16 \alpha_5 \alpha_6 + 18 
\alpha_6 \alpha_6)
\label{eq:sig_ijlm_jklm}
\end{align}
Terms such as $\alpha_1 \alpha_1$, $\alpha_1 \alpha_2$ and others give rise to wavenumber 
integration of the form $\int d\bs{\kappa}d\bs{\kappa'}~
e^{-\mi (\boldsymbol{\kappa} + \boldsymbol{\kappa'}) \bcdot \boldsymbol{x}_g}\times(\cdots)$, which
upon substitution into \eqref{eq:int_rhs_dik} leads to time integrals of the following form.
\begin{align}
\int_{-\infty}^{t} \exp \left ( -\frac{t-t'}{T_{\epsilon \epsilon} } \right )~
e^{-\mi (\boldsymbol{\kappa}+\boldsymbol{\kappa}') \bcdot \boldsymbol{x}_g}~dt'
&= \frac{1} { \left ( \frac{1}{T_{\epsilon \epsilon}} \right )- 
\mi (\boldsymbol{\kappa}+\boldsymbol{\kappa}') \bcdot \boldsymbol{g}\tau_v } \notag \\
&= \frac{\left ( \frac{1}{T_{\epsilon \epsilon}} \right )+ 
\mi (\boldsymbol{\kappa}+\boldsymbol{\kappa}') \bcdot \boldsymbol{g}\tau_v } 
{ \left ( \frac{1}{T_{\epsilon \epsilon}} \right )^2 +  
[ (\boldsymbol{\kappa}+\boldsymbol{\kappa}') \bcdot \boldsymbol{g}\tau_v ]^2}
\label{eq:timeint_epseps}
\end{align}
It may be noted that in \eqref{eq:timeint_epseps}, the imaginary part on the RHS is 
odd in $\bs{g}$, whereas the drift flux is tensorially constrained to be 
even in $\bs{g}$.  Thus, the imaginary part does not contribute to the overall drift flux. 
Further details of the evaluation of the RHS of \eqref{eq:sig_ijlm_jklm} are presented 
in Appendix \ref{sec:app_a1a1}.

Next we evaluate the term $\langle \rho_{ij}(t) ~\rho_{lm}(t') \rangle ~
\langle \rho_{jk}(t) ~\rho_{lm}(t') \rangle$ in \eqref{eq:int_rhs_dik}. 
This again involves applying the appropriate tensorial constraints 
on each of the two correlations as follows. 
\begin{gather}
\langle \rho_{ij}(\bs{x},t) ~\rho_{lm}(\bs{x}',t') \rangle = 
\int d\boldsymbol{\kappa} ~ \langle \widehat{\rho}_{ij} (\bs{\kappa},t) ~
\widehat{\rho}^{*}_{lm} (\bs{\kappa},t)\rangle ~ 
e^{-\mi \boldsymbol{\kappa} \bcdot \boldsymbol{x}_g} = \mathscr{M}_{ijlm} = \notag \\ 
\beta_2(\delta_{im} \delta_{jl} - \delta_{il} \delta_{jm}) + 
\beta_6(\delta_{i3} \delta_{l3} \delta_{jm} - \delta_{i3} \delta_{m3} \delta_{jl} - %\notag \\
\delta_{j3} \delta_{l3} \delta_{im} + \delta_{j3} \delta_{m3} \delta_{il})
\label{eq:rho_ijlm}
\end{gather}
The criteria for determining $\beta$'s---provided in Appendix \ref{sec:app_tensor}---yield  
\begin{gather}
\beta_2 = \frac{1}{2} (2C_2 - C_1); \hspace{0.2in} \beta_6 = \frac{1}{2} (3C_2 - C_1)
\end{gather}
where
\begin{gather}
C_1 = \frac{\nu}{2 \langle \zeta \rangle} 
\left [ \frac{1}{\pi} \int d\boldsymbol{\kappa} ~E(\kappa) 
~e^{-\mi \boldsymbol{\kappa} \bcdot \boldsymbol{x}_g} \right ] \\
C_2 = \frac{\nu}{2 \langle \zeta \rangle} 
\left [ \int d\boldsymbol{\kappa} ~\kappa_j \kappa_j~\frac{E(\kappa) }{4\pi \kappa^2}
\left (1 + \frac{\kappa_3^2}{\kappa^2} \right )
e^{-\mi \boldsymbol{\kappa} \bcdot \boldsymbol{x}_g} \right ] 
\end{gather}
Thus, the product $\langle \rho_{ij}(t) ~\rho_{lm}(t') \rangle ~
\langle \rho_{jk}(t) ~\rho_{lm}(t') \rangle$ in \eqref{eq:int_rhs_dik} 
can now be written as 
\begin{gather}
\langle \rho_{ij}(t) ~\rho_{lm}(t') \rangle ~
\langle \rho_{jk}(t) ~\rho_{lm}(t') \rangle = \delta_{ik}(-4 \beta_2 \beta_2 + 4 \beta_2 \beta_6 -
2 \beta_6 \beta_6) + \delta_{i3}\delta_{k3} (4 \beta_2 \beta_6 - 2 \beta_6 \beta_6)
\label{eq:rho_ij_jk_lm_beta}
\end{gather}
Terms on the RHS of \eqref{eq:rho_ij_jk_lm_beta} such as $\beta_2 \beta_2$, $\beta_2 \beta_6$ and 
$\beta_6 \beta_6$ contain wavenumber integration of the form $\int d\bs{\kappa}d\bs{\kappa'}~
e^{-\mi (\boldsymbol{\kappa} + \boldsymbol{\kappa'}) \bcdot \boldsymbol{x}_g}\times (\cdots)$, which 
upon substitution into \eqref{eq:int_rhs_dik} leads to a 
time integration similar to that in  \eqref{eq:timeint_epseps}, with the 
$T_{\epsilon \epsilon}$ replaced by $T_{\zeta \zeta}$.

Recalling the integral $\int_{-\infty}^{t} d_{ik}~dt'$ in \eqref{eq:int_rhs_dik}, 
we can evaluate terms such as 
\begin{align}
\int_{-\infty}^{t} dt'~
\left \langle \epsilon(t) ~\epsilon(t') \right \rangle \langle \sigma_{ij}(t) ~\sigma_{lm}(t') \rangle ~\langle \sigma_{jk}(t) ~\sigma_{lm}(t') \rangle
\end{align}
by applying the time integral in \eqref{eq:timeint_epseps} along with 
\eqref{eq:sig_ijlm}-\eqref{eq:sig_ijlm_jklm}.  
The final form of drift flux for DF2 is analogous to that in \eqref{eq:drift_flux_gamma_terms_final} 
and is given by
\begin{eqnarray}\label{eq:drift_flux_2}
q_i^d(\boldsymbol{r},t) = -\langle P\rangle(r,\theta)~ 2r_k ~
\frac{St_\eta^2}{\Gamma_\eta^2} ~[\lambda_1' \left (\delta_{ik} - \delta_{i3} \delta_{k3} \right ) + 
\lambda_2' \delta_{i3} \delta_{k3}]
\end{eqnarray} 
where $\lambda_1'$ and $\lambda_2'$ are the coefficients for DF2.
The expressions for $\lambda_1'$ and $\lambda_2'$ are extremely involved and are not explicitly  
presented.  In fact, \eqref{eq:sig_ijlm_jklm} gives rise to 
thirteen separate integrations of the general form shown in 
\eqref{eq:timeint_epseps}, while \eqref{eq:rho_ij_jk_lm_beta} gives rise to three more such integrals.  
Each of these integrals 
is evaluated through numerical quadrature, and then assembled using \eqref{eq:sig_ijlm_jklm} and 
\eqref{eq:rho_ij_jk_lm_beta} during runtime (of the computational code).

\subsection{Diffusion Flux}
\label{subsec:diff}
Applying \eqref{eq:w_i} in the diffusion flux given by \eqref{eq:diff_flux}, and 
retaining only the leading order term yields the following form 
of the diffusion flux \citep{chun2005}
\begin{align}
q_i^D(\boldsymbol{r}) 
%& \approx -\frac{\partial \langle P\rangle}{\partial r_j}~
%r_m r_n \int_{-\infty}^{t} \left \langle \Gamma_{im}(t) ~ 
%\Gamma_{jn}(t') \right \rangle ~ dt' \notag \\
 = -\mathscr{D}_{ij} \frac{\partial \langle P\rangle}{\partial r_j}
\label{eq:diff_flux_simplified}
\end{align}
with the diffusivity tensor 
\begin{align}
\mathscr{D}_{ij} = r_m r_n \int_{-\infty}^{t} \left \langle \Gamma_{im}(t) ~ 
\Gamma_{jn}(t') \right \rangle ~ dt' 
= r_m r_n ~Q_{imjn} \label{eq:a_imjn}
\end{align}
where $\Gamma_{im}(t) = \Gamma_{im}(\boldsymbol{x}(t),t)$, 
$\Gamma_{jn}(t') = \Gamma_{jn}(\boldsymbol{x}(t'),t')$.

In writing \eqref{eq:diff_flux_simplified}, we have 
invoked the assumption that the pair separation does not 
change appreciably over the correlation time for the ``seen" fluid velocity 
gradient.  Such an approximation has been referred to as the local diffusion analysis in 
the \citet{chun2005} study, and is particularly suitable for the case of 
rapidly settling particle pairs.  As noted by \citet{peter2015b}, gravity  
reduces the Lagrangian time scales of strain-rate and 
rotation-rate along the particle trajectories.  Therefore, in the rapidly settling 
limit, one would anticipate these time scales to be significantly smaller 
than those in the zero gravity case.  Thus, it is reasonable to assume 
the pair separation to be essentially constant in these reduced 
correlation times of the fluid velocity gradient. 
 
Analogous to the drift analysis, we can express the two-time correlation 
$\langle \Gamma_{im}(t) ~ \Gamma_{jn}(t') \rangle$ in terms of 
two-point Eulerian correlation as 
\begin{eqnarray}
\left \langle \Gamma_{im}(t) ~ \Gamma_{jn}(t') \right \rangle &=& 
\left \langle \Gamma_{im}(\boldsymbol{x};t) ~ \Gamma_{jn}[\boldsymbol{x}+\boldsymbol{g}\tau_v(t'-t);t] \right \rangle \nonumber \\
&=& \int d\boldsymbol{\kappa} ~\kappa_m \kappa_n ~ \Phi_{ij}(\boldsymbol{\kappa}) ~ 
e^{\mi \boldsymbol{\kappa} \bcdot \boldsymbol{g}\tau_v(t'-t)} \nonumber \\
&=& \int d\boldsymbol{\kappa} ~\kappa_m \kappa_n \frac{E(\kappa)}{4 \pi \kappa^2} 
\left (\delta_{ij} - \frac{\kappa_i \kappa_j}{\kappa^2} \right ) ~ 
e^{\mi \boldsymbol{\kappa} \bcdot \boldsymbol{g}\tau_v(t'-t)} 
\end{eqnarray}
Thus, the diffusivity tensor may be written as
\begin{eqnarray}
\mathscr{D}_{ij}(\boldsymbol{r}) &=& r_m r_n \int d\boldsymbol{\kappa}  ~\kappa_m \kappa_n  
\frac{E(\kappa)}{4 \pi \kappa^2} 
\left (\delta_{ij} - \frac{\kappa_i \kappa_j}{\kappa^2} \right ) ~ 
\int_{-\infty}^{0} e^{\mi \boldsymbol{\kappa} \bcdot \boldsymbol{g}\tau_v t} ~ dt \nonumber \\
&=& \frac{1}{2} r_m r_n \int d\boldsymbol{\kappa} ~\kappa_m \kappa_n  \frac{E(\kappa)}{4 \pi \kappa^2} 
\left (\delta_{ij} - \frac{\kappa_i \kappa_j}{\kappa^2} \right ) ~ 
\delta(\boldsymbol{\kappa} \bcdot \boldsymbol{g}\tau_v) \nonumber \\
&=& r_m r_n \frac{1}{2{g}\tau_v} \int d\boldsymbol{\xi} ~\xi_m \xi_n \frac{E(\xi)}{4 \pi \xi^2} 
\left (\delta_{ij} - \frac{\xi_i \xi_j}{\xi^2} \right ) \label{eq:dij_spectrum}
%&=& r_m r_n A_{imjn}
\end{eqnarray} 
where $\boldsymbol{\xi} = (\xi_1,\xi_2,0)$ is the wavenumber vector in 
the homogeneous $x_1-x_2$ plane.

Using \eqref{eq:a_imjn} and \eqref{eq:dij_spectrum}, and applying the tensor constraints 
on the fourth order tensor $Q_{imjn}$ yields (details of the tensor analysis are in Appendix \ref{sec:app})
\begin{align}
&Q_{imjn} = \frac{1}{2{g}\tau_v} \int d\boldsymbol{\xi} ~\xi_m \xi_n  \frac{E(\xi)}{4 \pi \xi^2} 
\left (\delta_{ij} - \frac{\xi_i \xi_j}{\xi^2} \right ) 
= \lambda_5 \left (\delta_{i3} \delta_{m3} \delta_{j3} \delta_{n3} - \delta_{i3} \delta_{j3} \delta_{mn} \right ) + 
\nonumber \\
&\lambda_6 \left (\delta_{in} \delta_{mj} + \delta_{im} \delta_{jn} - 3 \delta_{ij} \delta_{mn} - 
\delta_{i3} \delta_{n3} \delta_{mj} - \delta_{j3} \delta_{n3} \delta_{im} - \delta_{i3} \delta_{m3} \delta_{jn} - 
\delta_{m3} \delta_{j3} \delta_{in} + \right. \nonumber \\ 
&\left . 2 \delta_{i3} \delta_{j3} \delta_{mn} + 3 \delta_{m3} \delta_{n3} \delta_{im}
\right )
\label{eq:lambda56}
\end{align}
which gives
\begin{eqnarray}
\lambda_5 &=& -\frac{3\pi}{16{g}\tau_v} \int_{\xi = 0}^{\infty} \xi~ E(\xi)~d{\xi}, \\
\lambda_6 &=& \frac{\lambda_5}{3}. \label{eq:lamb6} 
\end{eqnarray}
Therefore,
\begin{eqnarray}
\mathscr{D}_{ij}(\boldsymbol{r}) &=& r_m r_n ~A_{imjn} \nonumber \\
&=& \lambda_6 \left [ (3r_3^2 - r^2) \delta_{i3} \delta_{j3} + 
2 r_i r_j + 3 (r_3^2 - r^2) \delta_{ij} - 2r_3 (r_j \delta_{i3} + r_i \delta_{j3})
\right ] \label{eq:dij_final}
\end{eqnarray}
%%%%
%It will be seen below that the radial component, $\mathscr{D}_{rr}$, of the 
%diffusivity tensor scales as $\lambda_6 r^2$.  Thus, a good estimate 
%of the time over which the PDF $\langle P \rangle$ evolves may be obtained 
%using the coefficient $\lambda_6$ which has the dimensions of inverse time.  
%To calculate $\lambda_6$ from \eqref{eq:lamb6}, we need the energy spectrum 
%$E(\kappa)$. The need for DNS-computed $E(\kappa)$ 
%may be obviated by using the following dimensionless form of $E(\kappa)$, valid
%in the limit $Re_\lambda \rightarrow \infty$ \citep{pope_2000}.
%\begin{gather}
%\frac{E(\kappa \eta)}{\eta u_\eta^2} = 1.5 f_\eta(\kappa \eta) \label{eq:enspec} \\
%f_\eta(\kappa \eta) = \mathrm{exp}\left \{-5.2\left ( [(\kappa \eta)^4 + c_\eta^4]^{1/4} -c_\eta\right ) \right \} \label{eq:feta}
%\end{gather}
%where $c_\eta \approx 0.4$ for $Re_\lambda \rightarrow \infty$, and $\eta$ and $u_\eta$ 
%are the Kolmogorov length and velocity scales.  The 
%integral in \eqref{eq:lamb6} is then evaluated through numerical quadrature.
%The characteristic time scale of $\langle P \rangle$ is thus obtained 
%to be $\approx 20.4132 \times (St_\eta/Fr) \times \tau_\eta \gg \eta/g \tau_v$.
%%%%
Having derived closures for the drift and diffusion fluxes, we present the 
analytical solution to the PDF equation \eqref{Eqn:PDF}.

\section{Solution of the PDF Equation}
\label{sec:solution}
We will solve the PDF equation \eqref{Eqn:PDF} in spherical coordinates.  At 
steady state, the governing equation for $\langle P \rangle (r,\theta)$ is given by 
\begin{eqnarray}
\frac{1}{r^2}\frac{\partial}{\partial r}\left(r^2q_r\right)+
\frac{1}{r\sin\!\theta}\frac{\partial}{\partial\theta}\left(\sin\!\theta q_{\theta}\right)=0
\label{eq:solmthd1}
\end{eqnarray}
where $q_r$ and $q_\theta$ are fluxes along the radial and polar directions.  These contain 
both the drift and diffusion fluxes, and are given by 
\begin{eqnarray}
q_r&=&v_r \langle P \rangle - \mathscr{D}_{rr}\frac{\partial \langle P \rangle}{\partial r}-\mathscr{D}_{r\theta}\frac{1}{r}\frac{\partial \langle P \rangle}{\partial\theta}
\nonumber
\\
q_{\theta}&=&v_{\theta} \langle P \rangle - \mathscr{D}_{r\theta}\frac{\partial \langle P \rangle}{\partial r}-\mathscr{D}_{\theta\theta}\frac{1}{r}\frac{\partial \langle P \rangle}{\partial\theta}
\nonumber
\\
v_r&=&-2r\left(\lambda_1\sin^2\theta+\lambda_2\cos^2\theta\right)St_\eta^2
\nonumber
\\
v_{\theta}&=&-2r\left(\lambda_1-\lambda_2\right)St_\eta^2\sin\!\theta\cos\!\theta
\nonumber
\\
\mathscr{D}_{rr}&=&\lambda_6 r^2\left(3\sin^4\!\theta-4\sin^2\!\theta\right)
\nonumber
\\
\mathscr{D}_{r\theta}&=&3\lambda_6 r^2\sin^3\!\theta\cos\!\theta
\nonumber
\\
\mathscr{D}_{\theta\theta}&=&-\lambda_6 r^2\left(\sin^2\!\theta+3\sin^4\!\theta\right)
\nonumber
\end{eqnarray}
The coefficients $\lambda_1$, $\lambda_2$ and $\lambda_6$ in the above equations 
are given in \eqref{eq:lamb1}, \eqref{eq:lamb2} and \eqref{eq:lamb6} respectively, while 
$\mathscr{D}_{rr}$, $\mathscr{D}_{r\theta}$ and $\mathscr{D}_{\theta\theta}$ are the components 
in spherical coordinates of the diffusivity tensor $\mathscr{D}_{ij}(\boldsymbol{r})$ 
in \eqref{eq:dij_final}.   When applying DF2, we use $\lambda_1'$ and $\lambda_2'$ in 
place of $\lambda_1$ and $\lambda_2$.

It is evident from the $q_r$ and $q_\theta$ equations that the variables $r$ and $\theta$ are separable.  
Also, the form of the PDF equation \eqref{eq:solmthd1} suggests a solution with a power law 
dependence on separation $r$.  Accordingly, we write 
$\langle P \rangle(r,\theta)=r^{\beta}f(\theta)$ and substitute this form into 
\eqref{eq:solmthd1}.  A change of variable $\mu=\cos\!\theta$ leads to
the following equation for $f(\mu)$ 
\begin{eqnarray}
a(\mu)\frac{d^2f}{d\mu^2}+b(\mu)\frac{df}{d\mu}+c(\mu)f=0
\label{eq:solmthd2}
\end{eqnarray}
where
\begin{eqnarray}
a(\mu)&=&\lambda_6(3\mu^2-4)(1-\mu^2)^2 
\nonumber
\\
b(\mu)&=&2(\lambda_2-\lambda_1)St^2\mu(1-\mu^2)-3\lambda_6\beta\mu(1-\mu^2)^2-
\nonumber
\\
       &&\lambda_6\mu(18\mu^2-22)(1-\mu^2)-3\lambda_6(\beta+3)\mu(1-\mu^2)^2
\nonumber
\\
c(\mu)&=&2(\lambda_2-\lambda_1)St^2(1-3\mu^2)-3\lambda_6\beta(1-\mu^2)(1-5\mu^2)+
\nonumber
\\
       &&(\beta+3)\left\{2\left[\lambda_1(1-\mu^2)+\lambda_2\mu^2\right]St^2-\lambda_6\beta(1+3\mu^2)(1-\mu^2)\right\}      
\nonumber
\end{eqnarray}
\subsection{Power Law Exponent $\beta$}
To find the power law exponent, we apply the constraint that at steady state, 
the net radial flux through a spherical surface of radius $r$ is zero, given by
\begin{equation}
\int_{-1}^{1}q_rd\mu=0
\end{equation}
leading to 
\begin{eqnarray}
\beta=\frac{\int_{-1}^{1}\left(A_rf(\mu)+B_{r\theta}\sqrt{1-\mu^2}\frac{df}{d\mu}\right)d\mu}
           {\int_{-1}^{1}B_{rr}f(\mu)d\mu}
\label{eq:solmthd3}
\end{eqnarray}
where
\begin{eqnarray}
A_r&=&-2\left[\lambda_1(1-\mu^2)+\lambda_2\mu^2\right]St_\eta^2 \nonumber \\
B_{rr}&=&-\lambda_6(1-\mu^2)(1+3\mu^2) \nonumber \\
B_{r\theta}&=&3\lambda_6\mu(1-\mu^2)\sqrt{1-\mu^2}. \nonumber
\end{eqnarray}
Since the drift flux scales as $St_\eta^2$, we seek $\beta=\beta_2~St_\eta^2$ ($\beta_2 > 0$), which 
then means that the numerator of \eqref{eq:solmthd3}, 
$\int_{-1}^{1}\left(A_rf(\mu)+B_{r\theta}\sqrt{1-\mu^2}\frac{df}{d\mu}\right)d\mu$, should 
also scale as $St_\eta^2$.  With these arguments, 
we seek a perturbation solution to \eqref{eq:solmthd2} of the form 
\begin{equation}
f(\mu)=f_0(\mu)+St_\eta^2 ~f_2(\mu).
\end{equation}

\subsection{Perturbation Solution for $f(\mu)$}
Substitution of 
$f(\mu)=f_0(\mu)+St_\eta^2 ~f_2(\mu)$ into \eqref{eq:solmthd2} and gathering terms 
that are $O(St^0)$ gives 
\begin{eqnarray}
a_0(\mu)\frac{d^2f_0}{d\mu^2}+b_0(\mu)\frac{df_0}{d\mu}=0 \label{eq:f0_eqn}
\end{eqnarray}
where
\begin{eqnarray}
a_0(\mu)&=&\lambda_6(3\mu^2-4)(1-\mu^2)^2 
\nonumber \\
b_0(\mu)&=&-\lambda_6\mu(18\mu^2-22)(1-\mu^2)-9\lambda_6\mu(1-\mu^2)^2
\nonumber
\end{eqnarray}
Equation \eqref{eq:f0_eqn} can be integrated to give
\begin{eqnarray}
\frac{df_0}{d\mu}=k_1\frac{(4-3\mu^2)^{1/2}}{(1-\mu^2)^2} \nonumber
\end{eqnarray}
which upon further integration leads to 
\begin{eqnarray}
f_0(\mu)=k_2+k_1 \left [ \frac{1}{2} \frac{\mu(4-3\mu^2)^{1/2}}{(1-\mu^2)} 
+2\tanh^{-1} \mu +\ln\left( \frac{4-3\mu+\sqrt{4-3\mu^2}}{4+3\mu+\sqrt{4-3\mu^2}}\right ) 
\right ]\nonumber
\end{eqnarray}
Recalling that $\mu = \cos \theta \in [-1,1]$, it can be seen that 
$f_0 \rightarrow \infty$ as $\mu \rightarrow \pm 1$.  These singularities 
prevent the normalization of the probability density $f_0$, suggesting that 
the integration constant $k_1 = 0$. 
%If $k_1\neq 0$ then $f_0\sim\frac{1}{1-\mu}$ as $\mu\rightarrow 1$ and is not normalizable.
%Further if $k_1\neq 0$ then $\frac{df_0}{d\mu}\neq 0$ at $\mu=0$ which violates symmetry.
Hence, we have $f_0(\mu)=k_2$. Using the normalization constraint 
$\int_0^1f_0d\mu=\frac{1}{4\pi}$ leads to $f_0=\frac{1}{4\pi}$.
%\begin{eqnarray}
%P(\mu)&=&P_0(\mu) 
%\nonumber \\
%Q(\mu)&=&Q_0(\mu)+St^2Q_2(\mu) 
%\nonumber \\
%R(\mu)&=&St^2R_2(\mu) 
%\nonumber 
%\end{eqnarray}

Having determined $f_0$, we now gather terms 
that are $O(St^2)$ as well as use $f_0=\frac{1}{4\pi}$, giving us
\begin{eqnarray}
a_0(\mu)\frac{d^2f_2}{d\mu^2}+b_0(\mu)\frac{df_2}{d\mu}+\frac{c_2(\mu)}{4\pi}=0 \nonumber
\end{eqnarray}
where
\begin{eqnarray}
%P_0(\mu)&=&\alpha_2(3\mu^2-4)(1-\mu^2)^2 
%\nonumber \\
%Q_0(\mu)&=&-\alpha_2\mu(18\mu^2-22)(1-\mu^2)-9\alpha_2\mu(1-\mu^2)^2
%\nonumber \\
%Q_2(\mu)&=&2(\lambda_2-\lambda_1)\mu(1-\mu^2)-6\alpha_2\beta_2\mu(1-\mu^2)^2
%\nonumber \\
c_2(\mu)&=&2(\lambda_2-\lambda_1)(1-3\mu^2)-3\lambda_6\beta_2(1-\mu^2)(1-5\mu^2)+
\nonumber
\\
       &&6\left[\lambda_1(1-\mu^2)+\lambda_2\mu^2\right]-3\lambda_6\beta_2(1+3\mu^2)(1-\mu^2)      
\label{eq:f2}
\end{eqnarray}
Equation \eqref{eq:f2} is a linear, inhomogeneous first order 
ordinary differential equation in $\frac{df_2}{d\mu}$, and can be integrated
using the integrating factor 
\begin{eqnarray}
{\mbox{Integrating Factor}}\,\, I=\exp\left[\int\frac{Q_0(\mu)}{P_0(\mu)}d\mu\right]=\frac{(1-\mu^2)^2}{(4-3\mu^2)^{1/2}} \nonumber
\end{eqnarray}
Thus, we have
\begin{eqnarray}
I\frac{df_2}{d\mu}=\int I\frac{-R_2(\mu)}{P_0(\mu)4\pi}d\mu+k_3 \label{eq:df2_dphi}
\end{eqnarray}
where $k_3$ is a constant of integration. 
%\begin{eqnarray}
%\frac{df_2}{d\mu}&=&\frac{1}{I}\int I\frac{-R_2(\mu)}{P_0(\mu)4\pi}d\mu+\frac{k_3}{I} \nonumber
%\\
%&=&\frac{(4-3\mu^2)^{1/2}}{(1-\mu^2)^2}\int \frac{(1-\mu^2)^2}{(4-3\mu^2)^{1/2}}\frac{-R_2(\mu)}{P_0(\mu)4\pi}d\mu+
%k_3\frac{(4-3\mu^2)^{1/2}}{(1-\mu^2)^2} \nonumber
%\\
%&=&\frac{1}{4\pi}\frac{\mu\left[\lambda_2+2\lambda_1+\alpha_2\beta_2(-3+2\mu^2)\right]}{2\alpha_2(1-\mu^2)^2}+
%k_3\frac{(4-3\mu^2)^{1/2}}{(1-\mu^2)^2} \nonumber
%\end{eqnarray}
To find $k_3$, we enforce symmetry $\frac{df}{d\mu}=0$ at $\mu=0$. Since the first term 
on the RHS of \eqref{eq:df2_dphi} is zero at $\mu=0$, it follows that $k_3=0$
in order to satisfy the symmetry requirement. Thus  
\begin{eqnarray}
\frac{df_2}{d\mu}=
\frac{1}{4\pi}\frac{\mu\left[\lambda_2+2\lambda_1+\lambda_6\beta_2(-3+2\mu^2)\right]}{2\lambda_6(1-\mu^2)^2}
\label{eq:solmthd4}
\end{eqnarray}

Referring to \eqref{eq:solmthd3}, in order for $\beta$ to scale as $St_\eta^2$, we use $f(\mu) = f_0 = 1/4\pi$ 
and $df/d\mu = df_2/d\mu$ in the numerator of \eqref{eq:solmthd3}, giving us
\begin{eqnarray}
\beta_2=\frac{\int_{0}^{1}\left(\frac{a_{r}}{4\pi}+b_{r\theta}\frac{df_2}{d\mu}\right)d\mu }
           {\int_{0}^{1}\frac{B_{rr}}{4\pi}d\mu}
\label{eq:solmthd5}
\end{eqnarray}
where 
\begin{eqnarray}
a_r&=&-2\left[\lambda_1(1-\mu^2)+\lambda_2\mu^2\right] \nonumber \\
B_{rr}&=&-\lambda_6(1-\mu^2)(1+3\mu^2) \nonumber \\
b_{r\theta}&=&3\lambda_6\mu(1-\mu^2)^2. \nonumber
\end{eqnarray}
Substitution of \eqref{eq:solmthd4} into \eqref{eq:solmthd5} and simplification thereafter leads to
\begin{eqnarray}
\beta_2=\frac{\lambda_2+2\lambda_1}{\lambda_6}
\label{eq:solmthd6}.
\end{eqnarray}
It may noted that $\beta_2 < 0$ as both $\lambda_1$ and $\lambda_2$ are $< 0$.

Using the above form of $\beta_2$ in \eqref{eq:solmthd4} we get 
\begin{eqnarray}
\frac{df_2}{d\mu}=-
\frac{1}{4\pi}\frac{\mu\beta_2}{(1-\mu^2)} \nonumber
\end{eqnarray}
which leads to
\begin{eqnarray}
f_2=\frac{\beta_2}{8\pi}\,\,\ln(1-\mu^2) + k_4 \nonumber
\end{eqnarray}
The unknown constant $k_4$ may be determined using $\int_{0}^{1} f_2 d\mu = 0$, yielding 
\begin{equation}
k_4 = -\frac{1}{4\pi} \beta_2 (\ln 2 - 1)
\end{equation}
Thus the complete solution for $f(\mu)$ is given by
\begin{eqnarray}
f(\mu)=\frac{1}{4\pi}\left[1+St_\eta^2 ~\beta_2  
\left (\frac{1}{2} \ln(1-\mu^2) - (\ln 2 - 1) \right ) \right].
\nonumber
\end{eqnarray}
%We normalize $f(\mu)$ such that $\frac{1}{K}\int_0^1 f(\mu)d\mu=\frac{1}{4\pi}$,
%yielding 
%\begin{eqnarray}
%K = 1 + St_\eta^2~\beta_2 (\ln 2 -1)
%\end{eqnarray}
Therefore $\langle P \rangle(r,\mu)$ is given by
\begin{eqnarray}
\langle P \rangle (r,\mu)=r^{\beta_2St^2} \frac{1}{4\pi}\left[1+St_\eta^2 ~\beta_2  
\left (\frac{1}{2} \ln(1-\mu^2) - (\ln 2 - 1) \right ) \right]
\label{eq:pdf_sol}
\end{eqnarray}
where $\beta_2$ is given by \eqref{eq:solmthd6}.  
%It is clear from 
%\eqref{eq:pdf_sol} that the PDF $\langle P \rangle(r,\theta)$ is minimum 
%at $\theta = 0$ and achieves a maximum as $\theta \rightarrow \pm \pi$.  

\section{Results}
\label{sec:results}
\subsection{Discussion of the PDF Solution}
\label{subsec:disc}

The PDF solution $\langle P \rangle(r,\mu)$ in \eqref{eq:pdf_sol} quantifies the 
dependence of particle clustering on separation $r$ and  
direction cosine $\mu$ ($=\cos\theta$), the latter quantifying anisotropy due to particle settling.
In the DNS by \citet{peter2015b}, they referred to $\langle P \rangle$ as the 
angular distribution function (ADF) $g(\bs{r})$, and expressed 
it in terms of the Legendre spherical harmonic functions, as below.
\begin{equation}
\frac{g(\bs{r})}{g(r)} = \sum_{l = 1}^{\infty} \frac{\mathscr{C}_{2l}^0(r)}{\mathscr{C}_0^0(r)}
Y_{2l}^0(\cos\theta)
\label{eq:gr_ratio}
\end{equation}
where 
\begin{equation}
g(r) = \mathscr{C}_0^0(r) = \int_{0}^{\pi} d\theta~ \sin\theta~ g(\bs{r})
\end{equation}
Applying the orthogonality of Legendre polynomials to \eqref{eq:gr_ratio}, we get
\begin{eqnarray}
\frac{\mathscr{C}_{2}^0(r)}{\mathscr{C}_0^0(r)} = \frac{5}{2} 
\frac{\int_{0}^{\pi} d\theta~ \sin\theta~ g(\bs{r})~Y_{2}^0(\cos\theta)} {g(r)} 
\end{eqnarray}
The corresponding value from the theory is
\begin{eqnarray}
\left [\frac{\mathscr{C}_{2}^0(r)}{\mathscr{C}_0^0(r)} \right ]_{\rm theory} 
%&=& \frac{5}{2} 
%\frac{\int_{0}^{\pi} d\theta~ \sin\theta~ \langle P \rangle(r,\theta)~Y_{2}^0(\cos\theta)} 
%{\int_{0}^{\pi} d\theta~ \sin\theta~ \langle P \rangle(r,\theta)} \notag \\
= \frac{5\beta_2 St_\eta^2}{12}
\label{eq:c2ratio_theory}
\end{eqnarray}
\citet{peter2015b} plotted the ratio $\mathscr{C}_{2}^0(r)/\mathscr{C}_0^0(r)$ as a function 
of $r$ for various $St_\eta > 0.3$.  These curves show that for $r \ll \eta$, the 
coefficient ratio 
becomes independent of $r$, suggesting that both $g(\bs{r})$ and $g(r)$ have the same 
functionality in $r$ for sub-Kolmogorov separations.  This was particularly the case 
for lower Stokes numbers.  The current theory also predicts that for 
$r \ll \eta$, the coefficient ratio is independent of $r$.   However, we could not 
directly compare the DNS and theory values of the coefficient ratio, as the theory 
is applicable for $St_\eta \ll 1$ and the DNS values of \citet{peter2015b} 
were for $St_\eta > 0.3$. 
It is evident from \eqref{eq:c2ratio_theory} that anisotropy due to gravity is 
small for $St_\eta \ll 1$.  A similar trend is noticed in the DNS of \citet{peter2015b}.
%However, it may be noted that particle clustering itself 
%decreases with gravity.  

%%%%%
%\begin{figure}
%
%    \centering
%        \psfrag{a}{$St_\eta$}
%        \psfrag{b}{$|\beta_2| * St_\eta^2$}
%        \psfrag{c}{$\infty$}
%        \centering
%        \hspace{-0.0in}
%        \includegraphics[scale=0.4]{./Plot_Paper1.eps}
%    \caption{Comparison of the power-law exponent $\beta_2$ obtained from theory and DNS 
%    of \cite{peter2015b}. 
%    (a) Results obtained using the two drift flux closures are shown; (b) Result obtained 
%    using only the second drift flux closure is shown.}
%    \label{fig:beta2_comp}
%\end{figure}

\subsection{Time Scale of PDF $\langle P \rangle$}
\label{subsec:result_pdf}
We have seen in Section \ref{sec:solution} that the radial component, $\mathscr{D}_{rr}$, of the 
diffusivity tensor scales as $\lambda_6 r^2$.  Thus, a good estimate 
of the time over which the PDF $\langle P \rangle$ evolves may be obtained 
using the coefficient $\lambda_6$ which has the dimensions of inverse time.  
To calculate $\lambda_6$ from \eqref{eq:lamb6}, we need the energy spectrum 
$E(\kappa)$. A fully analytical and universal result 
may be obtained by using the following dimensionless form of $E(\kappa)$---valid 
in the limit $Re_\lambda \rightarrow \infty$---that 
follows from Kolmogorov's first similarity hypothesis \citep{pope_2000}.
\begin{gather}
\frac{E(\kappa \eta)}{\eta u_\eta^2} = 1.5~ (\kappa \eta)^{-5/3}~
f_\eta(\kappa \eta) \label{eq:enspec} \\
f_\eta(\kappa \eta) = \mathrm{exp}\left \{-5.2\left ( [(\kappa \eta)^4 + c_\eta^4]^{1/4} -c_\eta\right ) \right \} \label{eq:feta}
\end{gather}
where $c_\eta \approx 0.4$ for $Re_\lambda \rightarrow \infty$, and $\eta$ and $u_\eta$ 
are the Kolmogorov length and velocity scales.  The 
integral in \eqref{eq:lamb6} is then evaluated through numerical quadrature.
The characteristic time scale of $\langle P \rangle$ is thus obtained 
to be $\approx 1.43118 \times (St_\eta/Fr) \times \tau_\eta \gg \eta/g \tau_v$.
Thus, the PDF evolves over time scales that are much longer than the settling time 
of a pair through a Kolmogorov-scale eddy.

\subsection{Prediction of Clustering through Universal Scaling}
\label{subsec:result_pdf}
The first drift closure DF1, and the diffusion flux 
have the advantage that the only statistical input they require 
is the energy spectrum $E(\kappa)$.  In contrast, DF2 requires the correlation 
length scales of dissipation rate and enstrophy as well.  The spectrum in \eqref{eq:enspec} enables 
us to obtain universal values of the drift and diffusion fluxes.
To determine the power law exponent $\beta_2$ for the spatial clustering 
of particles, we first non-dimensionalize
the drift and diffusion fluxes using the Kolmogorov length and time scales. 
We then substitute \eqref{eq:enspec} into the integrals in 
\eqref{eq:lamb1} and \eqref{eq:lamb2} for $\lambda_1$ and $\lambda_2$ of DF1 and also 
in \eqref{eq:lambda56} and \eqref{eq:lamb6} for the diffusion flux.  Finally, the integrals are 
evaluated through numerical quadrature.   

\begin{figure}

    \centering

        \psfrag{a}{\Large{$St_\eta$}}
        \psfrag{b}[c]{\Large{$-\beta_2 \times St_\eta ^2$}}
        \psfrag{c}{$\infty$}
        \centering
        \hspace{-0.0in}
        \includegraphics[scale=0.5]{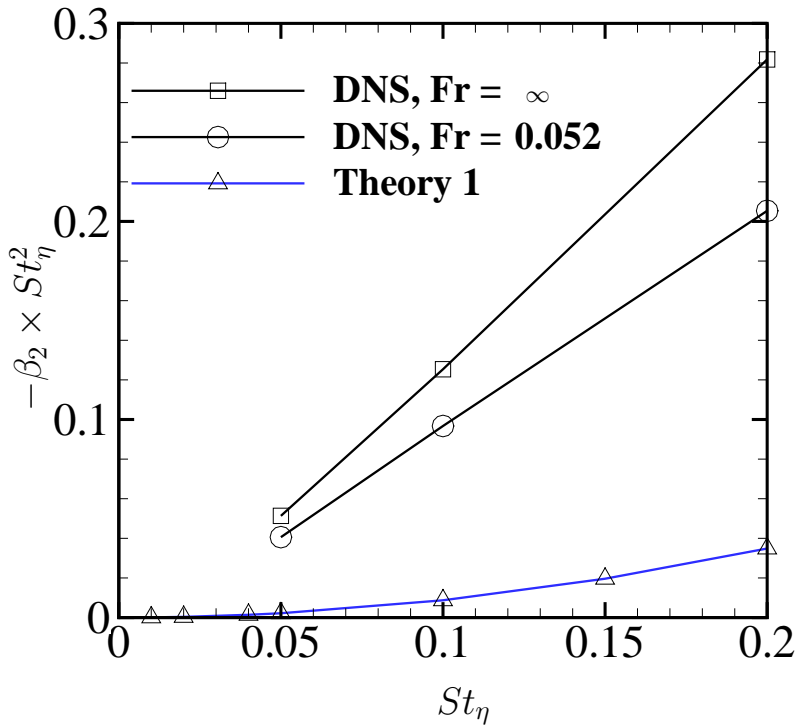}
 
    \caption{Power-law exponent $\beta_2$ obtained from DF1 in conjunction with 
    the universal energy spectrum (referred as Theory 1).  Also 
    shown are the DNS data of \citet{peter2015b} for $Fr = \infty$ and $Fr = 0.052$ at $Re_\lambda = 398$.}
    \label{fig:beta2}
\end{figure}

The $\beta_2$'s obtained using the above process are shown as a function 
of Stokes number in figure \ref{fig:beta2}.  Also shown are the DNS data from \citet{peter2015b} 
both with and without gravity ($Fr = 0.052$ and $Fr = \infty$, respectively)
at $Re_\lambda = 398$.   We see that the theory-predicted $\beta_2$'s 
are lower than the DNS values for both $Fr = 0.052$ and $Fr = \infty$.  
It may noted that the theory is derived for $Fr \ll 1$.  In addition, DF1 does not 
capture the two-time auto- and cross-correlations of strain-rate and rotation-rate invariants, which  
constitute the mechanisms responsible for particle clustering.

In the Part II paper, we present a direct comparison of theory 
predictions of particle clustering with {\it our} DNS data.  
Results obtained using both DF1 and DF2 will be presented. Turbulence 
and particle statistics needed as inputs to the theory will be obtained from the 
DNS runs.  The dependence of clustering on both separation and angular direction will be quantified.

\section{Conclusions} 
\label{sec:conclusions}
In Part I of this two-part study, we presented the derivation of closures  
for the drift and diffusion fluxes in the PDF equation for the 
pair relative positions $\bs{r}$.  The theory focuses on pair separations 
smaller than the Kolmogorov length scale, at which separations the theory 
approximates the fluid velocity field as being locally linear.  This allows 
us to express the fluid velocity differences between the secondary and primary particles 
of a pair in terms of the fluid velocity gradient at the location of the 
primary particle and their relative position.  Drift closures are obtained by expressing 
the pair relative velocity $W_i$ as a perturbation 
expansion in the Stokes number $St_\eta$.  

The drift flux contains the time integral of the third and fourth moments of the ``seen" fluid  
velocity gradients along the trajectories of primary particles. 
These moments are analytically resolved by making approximations  
regarding the ``seen" velocity gradient.  Accordingly, two 
closure forms, DF1 and DF2, are derived specifically for the drift flux. DF1 
is based on the assumption that the fluid velocity gradient ``seen" by the primary particle 
has a Gaussian distribution.  In DF2, we assume that 
the ``seen" strain-rate and rotation-rate tensors 
scaled by the dissipation rate and enstrophy, respectively, are 
normally distributed.  Unlike DF1, DF2 captures the two-time autocorrelations and cross-correlations 
of the strain-rate and rotation-rate invariants.  Time integrals of these 
correlations quantify the radially inward drift flux responsible for particle clustering.  
Analytical form  of the PDF $\langle P \rangle (r,\theta)$ is then obtained 
with a power-law dependence on separation~$r$.  Analogous to the theoretical result 
of \citet{chun2005} for non-settling pairs, and that of \citet{fouxon2015} for rapidly 
settling pairs, the power-law exponent scales as $St_\eta^2$.  The anisotropy in 
clustering due to gravity is also quantified by deriving an analytical expression for the 
ratio of coefficients in the spherical harmonics expansion of the PDF.  
As observed in the DNS of \citet{peter2015b}, when $St_\eta < 1$, the PDF obtained from the theory 
is only weakly anisotropic.  Predictions of particle clustering obtained from DF1 in conjunction with 
the universal Kolmogorov energy spectrum are presented, and compared with the DNS data 
of \citet{peter2015b}.  A more detailed and rigorous comparison of theory and DNS results  
is presented in the Part II paper.

\section*{Acknowledgements}

SLR and VKG gratefully acknowledge NSF support through the grant CBET-1436100.

\appendix 
\section{Tensorial Constraints}
\label{sec:app_tensor}
\subsection{$\langle \sigma_{ij}(t) ~\sigma_{lm}(t') \rangle = \mathscr{L}_{ijlm}$} 
Gravitational acceleration induces anisotropy along the $x_3$ direction, but homogeneity 
is satisfied along the $x_1$ and $x_2$ directions.  Accordingly, the fourth order tensor $A_{ijlm}$ 
in \eqref{eq:sig_ijlm} may be represented as 
\begin{align}
& \mathscr{L}_{ijlm} = \alpha_1 ~\delta_{ij} \delta_{lm} + \alpha_2 ~\delta_{im} \delta_{jl} + 
\alpha_3 ~\delta_{il} \delta_{jm} + \alpha_4 ~\delta_{i3} \delta_{j3} \delta_{l3} \delta_{m3} + 
\alpha_5 ~\delta_{i3} \delta_{j3} \delta_{lm} + \notag \\
& \alpha_6 ~\delta_{i3} \delta_{l3} \delta_{jm} + 
\alpha_7 ~\delta_{i3} \delta_{m3} \delta_{jl} + \alpha_8 ~\delta_{j3} \delta_{l3} \delta_{im} + 
\alpha_9 ~ \delta_{j3} \delta_{m3} \delta_{il} + \alpha_{10} ~ \delta_{l3} \delta_{m3} \delta_{ij}
\label{eq:alpha1_through_10}
\end{align}
where 
\begin{align}
\mathscr{L}_{ijlm} = \int d\boldsymbol{\kappa} ~ \langle \widehat{\sigma}_{ij} (\bs{\kappa},t) ~
\widehat{\sigma}^{*}_{lm} (\bs{\kappa},t)\rangle ~ 
e^{-\mi \boldsymbol{\kappa} \bcdot \boldsymbol{x}_g} 
\label{eq:A_ijlm}
\end{align}
Evaluation of the correlation $\langle \widehat{\sigma}_{ij} (\bs{\kappa},t) ~
\widehat{\sigma}^{*}_{lm} (\bs{\kappa},t)\rangle$ is presented in Appendix \ref{sec:app1}.
The coefficients $\alpha_1$ through $\alpha_{10}$ in \eqref{eq:alpha1_through_10} 
are determined using the following criteria. 
\begin{itemize}
\item Continuity: $\mathscr{L}_{iilm} =0$; $\mathscr{L}_{ijmm} = 0$
\item Symmetry: $\mathscr{L}_{ijlm} = \mathscr{L}_{ijml}$; $\mathscr{L}_{jilm} = \mathscr{L}_{lmij}$
\item Additional Independent Equations:
\begin{eqnarray*}
\mathscr{L}_{ijij} = B_1 \\
\mathscr{L}_{3333} = B_2 \\
\mathscr{L}_{3j3j} = B_3
\end{eqnarray*}
where 
\begin{gather}
B_1 = \frac{\nu}{2 \langle \epsilon \rangle} 
\left [ \frac{1}{\pi} \int d\boldsymbol{\kappa} ~E(\kappa) 
~e^{-\mi \boldsymbol{\kappa} \bcdot \boldsymbol{x}_g} \right ] 
\label{eq:b1} \\
B_2 = \frac{\nu}{2 \langle \epsilon \rangle} 
\left [ 4 \int d\boldsymbol{\kappa} ~\kappa_3^2~\frac{E(\kappa) }{4\pi \kappa^2}
\left (1 - \frac{\kappa_3^2}{\kappa^2} \right )
e^{-\mi \boldsymbol{\kappa} \bcdot \boldsymbol{x}_g} \right ] 
\label{eq:b2} \\
B_3 = \frac{\nu}{2 \langle \epsilon \rangle} 
\left [ \int d\boldsymbol{\kappa} ~\kappa_j \kappa_j~\frac{E(\kappa) }{4\pi \kappa^2}
\left (1 + \frac{\kappa_3^2}{\kappa^2} \right )
e^{-\mi \boldsymbol{\kappa} \bcdot \boldsymbol{x}_g} \right ] 
\label{eq:b3} 
\end{gather}
\end{itemize}
\subsection{$\langle \rho_{ij}(t) ~\rho_{lm}(t') \rangle = \mathscr{M}_{ijlm}$}
$\mathscr{M}_{ijlm}$ may be represented as 
\begin{align}
& \mathscr{M}_{ijlm} = \beta_1 ~\delta_{ij} \delta_{lm} + \beta_2 ~\delta_{im} \delta_{jl} + 
\beta_3 ~\delta_{il} \delta_{jm} + \beta_4 ~\delta_{i3} \delta_{j3} \delta_{l3} \delta_{m3} + 
\beta_5 ~\delta_{i3} \delta_{j3} \delta_{lm} + \notag \\
& \beta_6 ~\delta_{i3} \delta_{l3} \delta_{jm} + 
\beta_7 ~\delta_{i3} \delta_{m3} \delta_{jl} + \beta_8 ~\delta_{j3} \delta_{l3} \delta_{im} + 
\beta_9 ~ \delta_{j3} \delta_{m3} \delta_{il} + \beta_{10} ~ \delta_{l3} \delta_{m3} \delta_{ij}
\end{align}
where 
\begin{align}
\mathscr{M}_{ijlm} = \int d\boldsymbol{\kappa} ~ \langle \widehat{\rho}_{ij} (\bs{\kappa},t) ~
\widehat{\rho}^{*}_{lm} (\bs{\kappa},t)\rangle ~ 
e^{-\mi \boldsymbol{\kappa} \bcdot \boldsymbol{x}_g}  
\end{align}
The unknown $\beta$'s are determined using the following constraints.
\begin{itemize}
\item Continuity: $\mathscr{M}_{iilm} = 0$; $\mathscr{M}_{ijmm} = 0$
\item Symmetry: $\mathscr{M}_{ijlm} = \mathscr{M}_{lmij}$; $\mathscr{M}_{lmij} = -\mathscr{M}_{ijml}$
\item Additional Independent Equations:
\begin{eqnarray*}
\mathscr{M}_{ijij} = C_1 \\
\mathscr{M}_{3j3j} = C_2
\end{eqnarray*}
\end{itemize}
where
\begin{gather}
C_1 = \frac{\nu}{2 \langle \zeta \rangle} 
\left [ \frac{1}{\pi} \int d\boldsymbol{\kappa} ~E(\kappa) 
~e^{-\mi \boldsymbol{\kappa} \bcdot \boldsymbol{x}_g} \right ] \\
C_2 = \frac{\nu}{2 \langle \zeta \rangle} 
\left [ \int d\boldsymbol{\kappa} ~\kappa_j \kappa_j~\frac{E(\kappa) }{4\pi \kappa^2}
\left (1 + \frac{\kappa_3^2}{\kappa^2} \right )
e^{-\mi \boldsymbol{\kappa} \bcdot \boldsymbol{x}_g} \right ] 
\end{gather}
\section{Evaluation of $\langle \widehat{\sigma}_{ij} (\bs{\kappa},t) ~
\widehat{\sigma}^{*}_{lm} (\bs{\kappa},t)\rangle$} 
\label{sec:app2}

Using the normalization of the strain-rate tensor defined in \eqref{eq:gamma_ij_norm},  we can 
write $\widehat{\sigma}_{ij}$ in terms of the Fourier coefficients of the fluid velocity as 
\begin{equation}
\widehat{\sigma}_{ij} (\bs{\kappa},t) = \sqrt{\frac{2\nu}{\epsilon(t)}} ~\frac{1}{2} ~
[ \mi \kappa_j \widehat{u}_i (\bs{\kappa},t) + \mi \kappa_i \widehat{u}_j (\bs{\kappa},t) ]
\end{equation}
where $\mi = \sqrt{-1}$. We now have
\begin{eqnarray}
\langle \widehat{\sigma}_{ij} (\bs{\kappa},t) ~
\widehat{\sigma}^{*}_{lm} (\bs{\kappa},t)\rangle = 
-\frac{\nu}{2} \left \langle \frac{1}{\epsilon(t)} ~
(\mi \kappa_j \widehat{u}_i + \mi \kappa_i \widehat{u}_j)~
(\mi \kappa_m \widehat{u}_l^* + \mi \kappa_l \widehat{u}_m^*)
\right \rangle \notag \\
\approx \frac{\nu}{2 \langle \epsilon \rangle} \left [
\kappa_j \kappa_m \langle \widehat{u}_i \widehat{u}_l^* \rangle + 
\kappa_j \kappa_l \langle \widehat{u}_i \widehat{u}_m^* \rangle +
\kappa_i \kappa_m \langle \widehat{u}_j \widehat{u}_l^* \rangle +
\kappa_i \kappa_l \langle \widehat{u}_j \widehat{u}_m^* \rangle 
\right ] \notag \\
= \frac{\nu}{2 \langle \epsilon \rangle} \left [
\kappa_j \kappa_m \Phi_{il}(\bs{\kappa},t) + 
\kappa_j \kappa_l \Phi_{im}(\bs{\kappa},t) +
\kappa_i \kappa_m \Phi_{jl}(\bs{\kappa},t)  +
\kappa_i \kappa_l \Phi_{jm}(\bs{\kappa},t) 
\right ] 
\label{eq:sig_ij_lm_fourier}
\end{eqnarray}
where we have applied $\widehat{\sigma}^{*}_{lm} (\bs{\kappa},t) = 
\widehat{\sigma}_{lm} (-\bs{\kappa},t)$, and 
$\Phi_{il}(\bs{\kappa},t) = 
\langle \widehat{u}_i(\bs{\kappa},t) ~ \widehat{u}_l^*(\bs{\kappa},t) \rangle$ is 
the velocity spectrum tensor (see equation \eqref{eq:velspec}).  

The constraint $\mathscr{L}_{ijij} = B_1$ can now be obtained from \eqref{eq:sig_ij_lm_fourier} 
as follows.
\begin{gather}
B_1 = \int d\boldsymbol{\kappa} ~ \langle \widehat{\sigma}_{ij} (\bs{\kappa},t) ~
\widehat{\sigma}^{*}_{ij} (\bs{\kappa},t)\rangle ~ 
e^{-\mi \boldsymbol{\kappa} \bcdot \boldsymbol{x}_g} = \notag \\
\frac{\nu}{2 \langle \epsilon \rangle} 
\int d\boldsymbol{\kappa} ~ \left [
\kappa_j \kappa_j \Phi_{ii}(\bs{\kappa},t)  + 
\kappa_j \kappa_i \Phi_{ij}(\bs{\kappa},t)  +
\kappa_i \kappa_j \Phi_{ji}(\bs{\kappa},t)  +
\kappa_i \kappa_i \Phi_{jj}(\bs{\kappa},t) 
\right ] ~ e^{-\mi \boldsymbol{\kappa} \bcdot \boldsymbol{x}_g}
\label{eq:eq_b1_app}
\end{gather}
Using in \eqref{eq:eq_b1_app} the velocity spectrum tensor  
$\Phi_{ij}(\bs{\kappa},t)$ (see equation \eqref{eq:velspec}), 
it is relatively straightforward to show that $\kappa_j \kappa_i \Phi_{ij}(\bs{\kappa},t) = 0$, and 
the remaining terms together are equal to $B_1$ in \eqref{eq:b1}.  
The integrals contained in $B_2$ and $B_3$ (equations \eqref{eq:b2} and \eqref{eq:b3}) 
can be arrived at in a similar manner.

Analogous to \eqref{eq:sig_ij_lm_fourier}, we can also write
\begin{eqnarray}
&\langle \widehat{\sigma}_{jk} (\bs{\kappa},t) ~
\widehat{\sigma}^{*}_{lm} (\bs{\kappa},t)\rangle = \notag \\
&\frac{\nu}{2 \langle \epsilon \rangle} \left [
\kappa_k \kappa_m \langle \Phi_{jl}(\bs{\kappa},t) \rangle + 
\kappa_k \kappa_l \langle \Phi_{jm}(\bs{\kappa},t) \rangle +
\kappa_j \kappa_m \langle \Phi_{kl}(\bs{\kappa},t) \rangle +
\kappa_j \kappa_l \langle \Phi_{km}(\bs{\kappa},t) \rangle 
\right ] \\
%\end{eqnarray}
%%%%
%\begin{eqnarray}
&\langle \widehat{\rho}_{ij} (\bs{\kappa},t) ~
\widehat{\rho}^{*}_{lm} (\bs{\kappa},t)\rangle = \notag \\
&\frac{\nu}{2 \langle \epsilon \rangle} \left [
\kappa_j \kappa_m \langle \Phi_{il}(\bs{\kappa},t) \rangle - 
\kappa_j \kappa_l \langle \Phi_{im}(\bs{\kappa},t) \rangle -
\kappa_i \kappa_m \langle \Phi_{jl}(\bs{\kappa},t) \rangle +
\kappa_i \kappa_l \langle \Phi_{jm}(\bs{\kappa},t) \rangle 
\right ]  \\
%\end{eqnarray}
%%%%
%\begin{eqnarray}
&\langle \widehat{\rho}_{jk} (\bs{\kappa},t) ~
\widehat{\rho}^{*}_{lm} (\bs{\kappa},t)\rangle =  \notag \\
&\frac{\nu}{2 \langle \epsilon \rangle} \left [
\kappa_k \kappa_m \langle \Phi_{jl}(\bs{\kappa},t) \rangle - 
\kappa_k \kappa_l \langle \Phi_{jm}(\bs{\kappa},t) \rangle -
\kappa_j \kappa_m \langle \Phi_{kl}(\bs{\kappa},t) \rangle +
\kappa_j \kappa_l \langle \Phi_{km}(\bs{\kappa},t) \rangle 
\right ] 
\end{eqnarray}

\section{Evaluation of time integrals containing $\alpha_1 \alpha_1$, $\alpha_1 \alpha_2$, $\ldots$}
\label{sec:app_a1a1}
Reproducing \eqref{eq:int_rhs_dik}
\begin{gather}
\int_{-\infty}^{t} d_{ik}~dt' = 
\frac{1}{4\nu^2} \Bigl \{ ~\frac{1}{3} \delta_{ik}
\left [ \langle \epsilon^2 \rangle T_{\epsilon \epsilon} + 
\langle \epsilon \zeta \rangle T_{\epsilon \zeta} - 
\langle \zeta \epsilon \rangle T_{\zeta \epsilon} - 
\langle \zeta \zeta \rangle T_{\zeta \zeta} 
\right ] +  \notag \\
2 \langle \epsilon^2 \rangle \int_{-\infty}^{t} 
\exp \left ( -\frac{t-t'}{T_{\epsilon \epsilon} } \right ) \langle \sigma_{ij}(t) ~\sigma_{lm}(t') \rangle ~
\langle \sigma_{jk}(t) ~\sigma_{lm}(t') \rangle~dt' - \notag \\
2 \langle \zeta^2 \rangle \int_{-\infty}^{t} 
\exp \left ( -\frac{t-t'}{T_{\zeta \zeta} } \right ) \langle \rho_{ij}(t) ~\rho_{lm}(t') \rangle ~
\langle \rho_{jk}(t) ~\rho_{lm}(t') \rangle~dt'
~\Bigr \}
\label{eq:int_rhs_dik_reprod}
\end{gather}
the term $\int_{-\infty}^{t} 
\exp \left ( -\frac{t-t'}{T_{\epsilon \epsilon} } \right ) \langle \sigma_{ij}(t) ~\sigma_{lm}(t') \rangle ~
\langle \sigma_{jk}(t) ~\sigma_{lm}(t') \rangle~dt'$, in turn, contains integrals such as 
(see equations \eqref{eq:alpha1_through_10} and \eqref{eq:A_ijlm})
\begin{gather}
\int_{-\infty}^{t} 
\exp \left ( -\frac{t-t'}{T_{\epsilon \epsilon} } \right ) \alpha_1 \alpha_1 
\end{gather}
which leads to integrals of the form
\begin{gather} 
\int_{-\infty}^{t} \exp \left ( -\frac{t-t'}{T_{\epsilon \epsilon} } \right )~
e^{-\mi (\boldsymbol{\kappa}+\boldsymbol{\kappa}') \bcdot \boldsymbol{x}_g}~dt' 
\int d{\bs{\kappa}}~ d{\bs{\kappa'}} E(\kappa) E(\kappa') \\
=\int d{\bs{\kappa}}~ d{\bs{\kappa'}} E(\kappa) E(\kappa') 
\frac{\left ( \frac{1}{T_{\epsilon \epsilon}} \right ) } 
{ \left ( \frac{1}{T_{\epsilon \epsilon}} \right )^2 +  
[ (\boldsymbol{\kappa}+\boldsymbol{\kappa}') \bcdot \boldsymbol{g}\tau_v ]^2}
\end{gather}
The integral $\int d{\bs{\kappa}}~ d{\bs{\kappa'}} E(\kappa) E(\kappa') \times 
(\cdots)$ is then evaluated 
in spherical coordinates through numerical quadrature.

\section{Diffusion Flux Tensor Constraints} \label{sec:app}

The fourth order tensor $Q_{imjn}$ may be represented as 
\begin{align}
& Q_{imjn} = \alpha_1 ~\delta_{im} \delta_{jn} + \alpha_2 ~\delta_{in} \delta_{mj} + 
\alpha_3 ~\delta_{ij} \delta_{mn} + \alpha_4 ~\delta_{i3} \delta_{m3} \delta_{j3} \delta_{n3} + 
\alpha_5 ~\delta_{i3} \delta_{m3} \delta_{jn} + \notag \\
& \alpha_6 ~\delta_{i3} \delta_{j3} \delta_{mn} + 
\alpha_7 ~\delta_{i3} \delta_{n3} \delta_{mj} + \alpha_8 ~\delta_{in} \delta_{m3} \delta_{j3} + 
\alpha_9 ~\delta_{ij} \delta_{m3} \delta_{n3} + \alpha_{10} ~\delta_{im} \delta_{j3} \delta_{n3}
\end{align}
where the coefficients $\alpha_1$ through $\alpha_{10}$ are determined using   
the following criteria. 

\begin{itemize}\itemsep=4pt
\item Continuity: $Q_{iijn} = 0$, $Q_{imjj} = 0$, $Q_{iijj} = 0$
\item Symmetry: $Q_{imjn} = Q_{jmin}$, $Q_{imjn} = Q_{jnim}$
\item Additional Independent Equations:
\begin{align}
Q_{3m3m} = \frac{\pi}{2{g}\tau_v} \int_{\xi = 0}^{\infty} \xi~ E(\xi)~d{\xi} \notag \\
Q_{imim} = \frac{\pi}{{g}\tau_v} \int_{\xi = 0}^{\infty} \xi~ E(\xi)~d{\xi} \notag
\end{align}
\end{itemize}

\bibliographystyle{jfm}
\bibliography{refpaper2}

\begin{thebibliography}{24}
\expandafter\ifx\csname natexlab\endcsname\relax\def\natexlab#1{#1}\fi
\def\au#1{#1} \def\ed#1{#1} \def\yr#1{#1}\def\at#1{#1}\def\jt#1{\textit{#1}}
  \def\bt#1{#1}\def\bvol#1{\textbf{#1}} \def\vol#1{#1} \def\pg#1{#1}
  \def\publ#1{#1}\def\arxiv#1{#1}\def\org#1{#1}\def\st#1{\textit{#1}}

\bibitem[Ayala {\em et~al.\/}(2008{\natexlab{{\em a\/}}})Ayala, Rosa \&
  Wang]{ayala08b}
{\sc \au{Ayala, O.}, \au{Rosa, B.} \& \au{Wang, L.-P.}} \yr{2008{\natexlab{{\em
  a\/}}}}  \at{Effects of turbulence on the geometric collision rate of
  sedimenting droplets. part 2. theory and parameterization}.  \jt{New J.
  Phys.}  \bvol{10},  \pg{075016}.

\bibitem[Ayala {\em et~al.\/}(2008{\natexlab{{\em b\/}}})Ayala, Rosa, Wang \&
  Grabowski]{ayala08a}
{\sc \au{Ayala, Orlando}, \au{Rosa, Bogdan}, \au{Wang, Lian-Ping} \&
  \au{Grabowski, Wojciech~W}} \yr{2008{\natexlab{{\em b\/}}}}  \at{Effects of
  turbulence on the geometric collision rate of sedimenting droplets. part 1.
  results from direct numerical simulation}.  \jt{New Journal of Physics}
  \bvol{10}~(7),  \pg{075015}.

\bibitem[Bartlett(1966)]{bartlett1966}
{\sc \au{Bartlett, J.~T.}} \yr{1966}  \at{The growth of cloud droplets by
  coalescence}.  \jt{Quarterly Journal of the Royal Meteorological Society}
  \bvol{92}~(391),  \pg{93--104}.

\bibitem[Bragg \& Collins(2014{\natexlab{{\em a\/}}})]{bragg2014a}
{\sc \au{Bragg, Andrew~D} \& \au{Collins, Lance~R}} \yr{2014{\natexlab{{\em
  a\/}}}}  \at{New insights from comparing statistical theories for inertial
  particles in turbulence: I. spatial distribution of particles}.  \jt{New
  Journal of Physics}  \bvol{16}~(5),  \pg{055013}.

\bibitem[Bragg \& Collins(2014{\natexlab{{\em b\/}}})]{bragg2014b}
{\sc \au{Bragg, Andrew~D} \& \au{Collins, Lance~R}} \yr{2014{\natexlab{{\em
  b\/}}}}  \at{New insights from comparing statistical theories for inertial
  particles in turbulence: Ii. relative velocities}.  \jt{New Journal of
  Physics}  \bvol{16}~(5),  \pg{055014}.

\bibitem[Chun {\em et~al.\/}(2005)Chun, Koch, Rani, Ahluwalia \&
  Collins]{chun2005}
{\sc \au{Chun, Jaehun}, \au{Koch, Donald~L}, \au{Rani, Sarma~L}, \au{Ahluwalia,
  Aruj} \& \au{Collins, Lance~R}} \yr{2005}  \at{Clustering of aerosol
  particles in isotropic turbulence}.  \jt{Journal of Fluid Mechanics}
  \bvol{536},  \pg{219--251}.

\bibitem[Dhariwal {\em et~al.\/}(2017)Dhariwal, Rani \& Koch]{rani2017}
{\sc \au{Dhariwal, Rohit}, \au{Rani, Sarma~L} \& \au{Koch, Donald~L}} \yr{2017}
   \at{Stochastic theory and direct numerical simulations of the relative
  motion of high-inertia particle pairs in isotropic turbulence}.  \jt{Journal
  of Fluid Mechanics}  \bvol{813},  \pg{205--249}.

\bibitem[Druzhinin(1995)]{druzhinin}
{\sc \au{Druzhinin, O.~A.}} \yr{1995}  \at{Dynamics of concentration and
  vorticity modification in a cellular flow laden with solid heavy particles}.
  \jt{Phys. Fluids A}  \bvol{7},  \pg{2132--2142}.

\bibitem[Druzhinin \& Elghobashi(1999)]{druzhinin99}
{\sc \au{Druzhinin, O.~A.} \& \au{Elghobashi, S.}} \yr{1999}  \at{On the decay
  rate of isotropic turbulence laden with microparticles}.  \jt{Phys. Fluids}
  \bvol{11},  \pg{602--610}.

\bibitem[Eaton \& Fessler(1994)]{eaton1994}
{\sc \au{Eaton, J.~K.} \& \au{Fessler, J.~R.}} \yr{1994}  \at{Preferential
  concentration of particles by turbulence}.  \jt{Int. J. Multiphase Flow}
  \bvol{20},  \pg{169--209}.

\bibitem[Ferry {\em et~al.\/}(2003)Ferry, Rani \& Balachandar]{rani2}
{\sc \au{Ferry, J.}, \au{Rani, S.~L.} \& \au{Balachandar, S.}} \yr{2003}  \at{A
  locally implicit improvement of the equilibrium eulerian method}.  \jt{Int.
  J. Multiphase Flow}  \bvol{29},  \pg{869--891}.

\bibitem[Fouxon {\em et~al.\/}(2015)Fouxon, Park, Harduf \& Lee]{fouxon2015}
{\sc \au{Fouxon, Itzhak}, \au{Park, Yongnam}, \au{Harduf, Roei} \& \au{Lee,
  Changhoon}} \yr{2015}  \at{Inhomogeneous distribution of water droplets in
  cloud turbulence}.  \jt{Physical Review E}  \bvol{92}~(3),  \pg{033001}.

\bibitem[Ireland {\em et~al.\/}(2016)Ireland, Bragg \& Collins]{peter2015b}
{\sc \au{Ireland, Peter~J}, \au{Bragg, Andrew~D} \& \au{Collins, Lance~R}}
  \yr{2016}  \at{The effect of reynolds number on inertial particle dynamics in
  isotropic turbulence. part 2. simulations with gravitational effects}.
  \jt{Journal of Fluid Mechanics}  \bvol{796},  \pg{659--711}.

\bibitem[Maxey(1987)]{maxey1987}
{\sc \au{Maxey, MR}} \yr{1987}  \at{The gravitational settling of aerosol
  particles in homogeneous turbulence and random flow fields}.  \jt{Journal of
  Fluid Mechanics}  \bvol{174},  \pg{441--465}.

\bibitem[Parishani {\em et~al.\/}(2015)Parishani, Ayala, Rosa, Wang \&
  Grabowski]{parishani2015}
{\sc \au{Parishani, H}, \au{Ayala, O}, \au{Rosa, B}, \au{Wang, L-P} \&
  \au{Grabowski, WW}} \yr{2015}  \at{Effects of gravity on the acceleration and
  pair statistics of inertial particles in homogeneous isotropic turbulence}.
  \jt{Physics of Fluids}  \bvol{27}~(3),  \pg{033304}.

\bibitem[Pope(2000)]{pope_2000}
{\sc \au{Pope, S.~B.}} \yr{2000} {\em Turbulent Flows\/}.  \publ{New York:
  Cambridge University Press}.

\bibitem[Rani \& Balachandar(2003)]{rani1}
{\sc \au{Rani, S.~L.} \& \au{Balachandar, S.}} \yr{2003}  \at{Evaluation of the
  equilibrium eulerian approach for the evolution of particle concentration in
  isotropic turbulence}.  \jt{Int. J. Multiphase Flow}  \bvol{29},
  \pg{1793--1816}.

\bibitem[Rani \& Balachandar(2004)]{rani3}
{\sc \au{Rani, S.~L.} \& \au{Balachandar, S.}} \yr{2004}  \at{Preferential
  concentration of particles in isotropic turbulence: A comparison of the
  lagrangian and the equilibrium eulerian approaches}.  \jt{Powder Technology}
  \bvol{141},  \pg{109--118}.

\bibitem[Rani {\em et~al.\/}(2014)Rani, Dhariwal \& Koch]{rani2014}
{\sc \au{Rani, Sarma~L}, \au{Dhariwal, Rohit} \& \au{Koch, Donald~L}} \yr{2014}
   \at{A stochastic model for the relative motion of high stokes number
  particles in isotropic turbulence}.  \jt{Journal of Fluid Mechanics}
  \bvol{756},  \pg{870--902}.

\bibitem[Ray \& Collins(2011)]{ray2011}
{\sc \au{Ray, Baidurja} \& \au{Collins, Lance~R}} \yr{2011}  \at{Preferential
  concentration and relative velocity statistics of inertial particles in
  navier--stokes turbulence with and without filtering}.  \jt{Journal of Fluid
  Mechanics}  \bvol{680},  \pg{488--510}.

\bibitem[Reade \& Collins(2000)]{reade2000effect}
{\sc \au{Reade, Walter~C} \& \au{Collins, Lance~R}} \yr{2000}  \at{Effect of
  preferential concentration on turbulent collision rates}.  \jt{Physics of
  Fluids (1994-present)}  \bvol{12}~(10),  \pg{2530--2540}.

\bibitem[Squires \& Eaton(1991)]{squires1991}
{\sc \au{Squires, Kyle~D} \& \au{Eaton, John~K}} \yr{1991}  \at{Preferential
  concentration of particles by turbulence}.  \jt{Physics of Fluids A: Fluid
  Dynamics}  \bvol{3}~(5),  \pg{1169--1178}.

\bibitem[Zaichik \& Alipchenkov(2003)]{zaichik03}
{\sc \au{Zaichik, L.~I.} \& \au{Alipchenkov, V.~M.}} \yr{2003}  \at{Pair
  dispersion and preferential concentration of particles in isotropic
  turbulence}.  \jt{Phys. Fluids}  \bvol{15},  \pg{1776--1787}.

\bibitem[Zaichik \& Alipchenkov(2007)]{zaichik2007}
{\sc \au{Zaichik, Leonid~I} \& \au{Alipchenkov, Vladimir~M}} \yr{2007}
  \at{Refinement of the probability density function model for preferential
  concentration of aerosol particles in isotropic turbulence}.  \jt{Physics of
  Fluids (1994-present)}  \bvol{19}~(11),  \pg{113308}.

\end{thebibliography}

\end{document}